\documentclass[prd,nofootinbib,preprint,superscriptaddress,floatfix]{revtex4}
\pdfoutput=1
\usepackage{feynmp}
\usepackage{amsmath, amssymb, amsthm, graphicx, fancyhdr,epsfig, slashed, mathrsfs}
\usepackage{subfigure}
\usepackage{tikzsymbols}
\usepackage{natbib}
\usepackage{float}
\usepackage{lipsum}
\usepackage{mathtools} 
\usepackage{setspace}
\usepackage[normalem]{ulem}
\usepackage{tikz,xcolor,hyperref}
\hypersetup{
     colorlinks=true,
     citecolor    = blue,
     linkcolor= blue
}
\usepackage[capitalise]{cleveref}

\definecolor{lime}{HTML}{A6CE39}
\DeclareRobustCommand{\orcidicon}{
	\begin{tikzpicture}
	\draw[lime, fill=lime] (0,0) 
	circle [radius=0.2] 
	node[white] {{\fontfamily{qag}\selectfont \tiny ID}};
	\draw[white, fill=white] (-0.0625,0.095) 
	circle [radius=0.007];
	\end{tikzpicture}
	\hspace{-2mm}
}

\foreach \x in {A, ..., Z}{\expandafter\xdef\csname orcid\x\endcsname{\noexpand\href{https://orcid.org/\csname orcidauthor\x\endcsname}
			{\noexpand\orcidicon}}
}

\begin{document}
\title {Boosted Dark Matter Driven by Cosmic Rays \\ 
             and Diffuse Supernova Neutrinos}
\author{Dilip Kumar Ghosh\orcidA}
\email{tpdkg@iacs.res.in}
\affiliation{School of Physical Sciences, Indian Association for the Cultivation of Science,\\ 2A $\&$ 2B Raja S.C. Mullick Road, Kolkata 700032, India}

\author{Tushar Gupta\orcidB}
\email{tushar.gupta@helsinki.fi}
\affiliation{Department of Physics, and Helsinki Institute of Physics, University of Helsinki, P.O. Box 64, 00014 Helsinki, Finland}

\author{Matti Heikinheimo\orcidC}
\email{matti.heikinheimo@helsinki.fi}
\affiliation{Department of Physics, and Helsinki Institute of Physics, University of Helsinki, P.O. Box 64, 00014 Helsinki, Finland}

\author{Katri Huitu\orcidD}
\email{katri.huitu@helsinki.fi}
\affiliation{Department of Physics, and Helsinki Institute of Physics, University of Helsinki, P.O. Box 64, 00014 Helsinki, Finland}

\author{Sk Jeesun\orcidE}
\email{skjeesun48@gmail.com}
\affiliation{School of Physical Sciences, Indian Association for the Cultivation of Science,\\ 2A $\&$ 2B Raja S.C. Mullick Road, Kolkata 700032, India}

\begin{abstract}
Direct detection of light dark matter can be significantly enhanced by up-scattering of dark matter with energetic particles in the cosmic ambient. This boosted dark matter flux can reach kinetic energies up to tens of MeV, while the typical kinetic energies of GeV mass dark matter particles in the Milky Way halo are of the order of keV.
Dark matter boosted by energetic diffuse supernova background neutrinos can be detected only through nuclear or electron scattering in ground-based detectors requiring a non-zero interaction of dark matter with nucleon or electron, in addition to its interaction with neutrino.
However, in the presence of dark matter-nucleon (electron) interaction, the scattering of dark matter with cosmic rays is unavoidable.
Thus, we consider boosted dark matter resulting from diffuse supernova neutrinos as well as cosmic protons (electrons) considering both energy-dependent and energy-independent scattering cross-sections between dark matter and standard model particles. We explore this scenario in dark matter detectors such as XENONnT and neutrino detectors like Super-Kamiokande.
\end{abstract}
\maketitle
\section{Introduction}
The presence of a non-baryonic, non-luminous dark matter (DM) component in our universe has been suggested by numerous astrophysical and cosmological observations \cite{Zwicky:1933gu,Rubin:1970zza,Clowe:2006eq,Planck:2018vyg}.
The cold  DM component in the late universe is also crucially important in the large-scale structure formation \cite{Planck:2018vyg}.
Despite all the observations of DM based on its gravitational interaction, the particle nature of DM remains an intriguing puzzle for both theoretical and experimental particle physics. 
In general, DM production mechanisms in the early universe require sizeable interaction strength with standard model (SM) particles \cite{Lee:1977ua,Scherrer:1985zt,Jungman:1995df,Srednicki:1988ce,Arcadi:2017kky, Roszkowski:2017nbc,Bertone:2004pz}.
The possible connection between SM and DM has motivated the community to explore the signature of DM in various direct detection experiments 
\cite{Feng:2010gw,Lin:2019uvt,Billard:2021uyg}. 
As the solar system moves in the DM halo, the Earth is expected to encounter the DM flux, which 
may scatter the target particle in the underground detectors producing detectable signals.
Scientists are building increasingly sensitive detectors to capture these interactions and unravel the dark matter 
puzzle.
Examples of such experiments include XENONnT \cite{XENON:2019zpr,XENON:2022ltv,XENON:2023cxc}, LUX-ZEPLIN (LZ) \cite{LZ:2022lsv,LZ:2023poo,LZCollaboration:2024lux}, Panda X-II \cite{PandaX-II:2020oim}, LUX \cite{LUX:2016ggv}, DEAP-360 \cite{DEAP:2019yzn}, DarkSide-50 \cite{DarkSide:2018kuk}. Such experiments use liquid inert elements like xenon or argon and consider nucleons or electrons as target particles depending on specific DM scenarios.
These experiments have reached the sensitivity to probe cross-sections well below the electro-weak (EW) strength at DM mass $\sim\mathcal{O}(10-100)$\, GeV posing a strong question mark on the popular weakly interacting massive particle (WIMP) scenario \cite{Lin:2019uvt}.
Experiments like XENONnT\cite{XENON:2023cxc} and LZ \cite{LZ:2022lsv,LZCollaboration:2024lux} has excluded DM-nucleon cross-section $\sigma_{\chi n}\gtrsim 10^{-47}~{\rm cm}^2$ at masses around $\sim 30$ GeV.

Direct detection of light DM ($M_\chi \lesssim 1$ GeV) is challenging due to the small energy transfer in DM-nucleus scattering. 
For such a small energy transfer, the recoil energy of the nucleus can be so small that it falls below the threshold of many conventional detectors.
Most of the DM detectors are insensitive to energy deposits below $\mathcal{O}$(keV), and the typical lab-frame velocities of DM particles in the Milky Way (MW) halo are $\mathcal{O}(10^{-3})$ in natural units. 
This implies non-relativistic kinetic energy $\frac12 M_\chi v_\chi^2\sim 10^{-6}M_\chi$ so that DM particles below the GeV scale bound to the MW halo are invisible to these detectors.
However, for DM-electron scattering, the existing detectors are sensitive to DM mass $\sim\mathcal{O}$(MeV) 
due to comparatively smaller target mass \cite{XENON:2019zpr}.
Despite that, sub-MeV DM remains elusive to electron recoil experiments, requiring new strategies to search for low-mass DM.
While there are many active avenues of development for low-threshold detectors that would be sensitive to much smaller energy deposits (e.g. SuperCDMS \cite{SuperCDMS:2017mbc, SuperCDMS:2018mne}, CRESST \cite{CRESST:2019jnq}), another possibility for observing light DM in conventional direct detection experiments is the existence of a potential sub-population of DM particles not gravitationally bound to the MW halo, with speeds greatly exceeding those of the halo particles.
The presence of high-energy cosmic rays (CR) (protons, electrons, and light nuclei) can be extremely helpful in this regard \cite{Boschini:2017fxq}.
As was initially discussed in \cite{Yin:2018yjn,Bringmann:2018cvk}, a fraction of the DM population can be boosted to very high velocities due to DM scattering with energetic particles in the cosmic ambient.
These boosted DM (BDM) particles may have sufficient energy to generate recoil energies above the threshold \cite{Bringmann:2018cvk}.
Since then, several BDM scenarios have been extensively investigated in the literature \cite{Bringmann:2018cvk,Alvey:2019zaa,Ema:2018bih,Cappiello:2018hsu,Dent:2019krz,Ema:2020ulo,Jho:2020sku,Dent:2020syp,Bardhan:2022bdg,Maity:2022exk,Alvey:2022pad,CDEX:2022fig,Bell:2023sdq,Dutta:2024kuj}\footnote{Several other studies to probe light DM can be found in \cite{Jaeckel:2020oet,Herrera:2023nww}.}. Severalworks have also explored the BDM signature considering different interactions, like scalar or vector-mediated, for both nuclear recoil \cite{Dent:2019krz,Ema:2020ulo} and electron recoil \cite{Dent:2020syp,Bardhan:2022bdg}.
Since only a fraction of the total DM population gets boosted, the most stringent constraints on BDM come from large-size neutrino detectors \cite{Dent:2019krz,Ema:2020ulo,Dent:2020syp,Bardhan:2022bdg}.

Previous works have considered BDM due to CR protons \cite{Bringmann:2018cvk,Alvey:2019zaa,Maity:2022exk,Alvey:2022pad,CDEX:2022fig,Bell:2023sdq,Dutta:2024kuj}, cosmic electrons \cite{Ema:2018bih,Cappiello:2018hsu,Dent:2020syp,Bardhan:2022bdg,Jho:2020sku} and the diffuse supernova neutrino background (DSNB) \cite{Das:2021lcr,Ghosh:2021vkt,DeRomeri:2023ytt,Das:2024ghw} separately. 
DSNB is the predicted energetic diffuse neutrino flux from supernova core-collapse 
throughout the evolution of the universe. Although such DSNB has not yet been detected, it is predicted to be isotropic in all three neutrino flavours. 
DM with non-zero neutrino interaction may also lead to high-energy DM boosted by such DSNB.
However, to detect such DSNB-boosted DM, a non-zero interaction of DM with either nucleon or electron is necessary, as the detectors are sensitive to these two recoils only \cite{Das:2021lcr,Ghosh:2021vkt,DeRomeri:2023ytt,Das:2024ghw}. Importantly, a non-zero nucleon (electron) interaction of DM will also lead to DM up-scattering by CR, thus potentially changing the constraints.
In this paper, we present combined constraints due to each of these two sources (CR and DSNB), updated with the most recent results from the relevant direct detection experiments.
For this, we consider two scenarios of DM: $(i)$ electrophilic DM where DM interacts with electrons and neutrinos; and $(ii)$ nucleophilic DM where DM interacts with nucleons and neutrinos.
We then perform a model agnostic analysis of DM recoil signature in experiments like XENONnT and 
Super Kamiokande considering the DM is boosted by both CR and DSNB. Leveraging on non-observation of any signal events, we put the combined joint constraints on DM neutrino and DM nucleon (electron) cross-sections. The inclusion of a DSNB boost significantly enhances the DM flux, leading to 
more stringent constraints.
The direct detection rates strongly depend on the mass of the particle mediating the interaction between DM and SM. If the mediator mass ($m_{\rm med}$) is smaller than the transferred momentum ($q$), the associated form factor for non-relativistic cross-section becomes momentum suppressed, resulting in diminished interaction rate \cite{Bardhan:2022bdg,Bell:2023sdq}.  Thus the direct detection constraints on BDM  obtained for light mediators ($m_{\rm med}\ll q$) are weaker and lie within the existing constraints \cite{Bardhan:2022bdg}. For this reason, we restrict our analysis to heavy mediators ($m_{\rm med}\gg q$) only.
%
Furthermore, we discuss the effect of energy-dependent cross-sections, such as heavy scalar and vector mediators, which lead to stronger constraints.
The effect of energy-dependent cross-sections with combined CR and DSNB boosted DM has not been performed till today to the best of our knowledge. 
There are several UV complete models like $L_e-L_\tau$, $B-3L_{\mu}$ \cite{Ghosh:2024cxi} etc. as well as effective models \cite{Chang:2014tea,Blennow:2019fhy} where, such electrophilic or nucleophilic DM models can be constructed. 
Instead of confining our study to a particular UV complete model, we adopt a model-independent approach. This allows us to explore a broader range of possible DM scenarios, without being limited by the assumptions and predictions of any specific model. By focusing on the phenomenological implications of DM interactions, we aim to derive general constraints that apply to a wide variety of dark matter models.

We organize the paper as follows. In \cref{sec:BDM}, we discuss the calculations of CR and DSNB boosted DM flux. The possible recoil signatures in detectors have been analysed in \cref{sec:recoil}.
After showing the results in \cref{sec:result}, we conclude in \cref{sec:conc}.

\section{Boosted DM from cosmic rays}
\label{sec:BDM}
This section presents the key assumptions and formalism to obtain the BDM flux. 
The DM is assumed to be contained in the halo with a Maxwell-Boltzmann velocity distribution with non-relativistic velocity dispersion ($\sigma_{v_\chi}\sim 10^{-3}$).  
These DM particles ($\chi$) will gain much kinetic energy upon scattering with high energy ($\sim$ GeV) cosmic particles. 
Since the DM velocities are small, one may assume DM to be at rest initially while computing the scattering with high-energy CRs.

\subsection{Cosmic ray scattering}
The BDM flux from the collision with cosmic ray particles (either electron (CRe) or proton (CRp)) is given by \cite{Bringmann:2018cvk},
\begin{equation}
    \frac{d\Phi_\chi}{dT_\chi}= D_{\rm eff} \frac{\rho_\chi}{M_\chi} \int^{\infty}_{T_i^{\rm min}(T_\chi)} \frac{d\Phi_i^{\rm LIS}}{dT_i} \frac{d\sigma_{\chi i}}{dT_\chi} dT_i,
    \label{eq:dm_flux}
\end{equation} 
where $\rho_\chi=0.3~{\rm GeV/cm}^3$ is the local DM density, $M_\chi$ is the DM mass, $\sigma_{\chi i}$ denotes  the elastic DM-$i$ scattering cross-section where $i$ stands for $e$ (CRe) and $p$ (CRp). 
$D_{\rm eff}$ is the effective distance within which the effects of CR particles have been accounted \cite{Bringmann:2018cvk,Dent:2020syp,Bardhan:2022bdg}
\footnote{Since the DM density distribution in the Milky Way is well known, the contribution is considered up to a few kpc from the galactic center \cite{Catena:2009mf}.
Assuming a homogeneous CR flux and performing the line of sight integration up to 1 kpc (10 kpc) leads to $D_{\rm eff}=0.998$ kpc (8.02 kpc)\cite{Bringmann:2018cvk,DeRomeri:2023ytt}.}.   
Though there is uncertainty in $D_{\rm eff}$, for simplicity, we have taken $D_{\rm eff}=1$ kpc to obtain conservative limits \cite{Bardhan:2022bdg}.
Changing $D_{\rm eff}$ by a factor of few will not change the limits dramatically, as later we will see that the constraints on $\sigma_{\chi i}\propto \sqrt{D_{\rm eff}}$.
The kinetic energy of DM (CR particle $i$) is depicted by $T_\chi~(T_i)$. $T_i^{\rm min}$ is the minimum kinetic energy required to upscatter a non-relativistic DM to energy $T_\chi$ and is given by \cite{Bringmann:2018cvk,Bardhan:2022bdg},
\begin{equation}
    T_{i}^{\rm min}(T_\chi) =
    \left(\frac{T_\chi}{2}-m_i \right) \left[1 \pm \sqrt{1+\frac{2 T_\chi (m_i+M_\chi)^2}{M_\chi(2 m_i-T_\chi)^2}}~~\right] ,
    \label{eq:tmin}
\end{equation}
where $m_i= m_{p}(m_e)$, for collision with CRp (CRe). Here $(+)$ or $(-)$ sign is applicable when $T_\chi > 2m_i$ or $T_\chi < 2m_i$, respectively.
The remaining key input in \cref{eq:dm_flux} is the local interstellar flux (LIS) of the cosmic particles. 
The local CR flux is generally obtained in literature by fitting the observations in Voyager \cite{Cummings:2016pdr}, Fermi-Lat \cite{Fermi-LAT:2011baq}, PAMELA \cite{PAMELA:2011bbe}, AMS-02 \cite{AMS:2014gdf} \footnote{See ref.\cite{Boschini:2017fxq} for a detailed review on reproducing CR flux from measurements using propagation packages.}.
For CRe, we use the parametrization used in \cite{Bardhan:2022bdg} and the flux ($d\Phi_e/dT_e d\Omega$) is given by,
\begin{equation}
    \dfrac{d\Phi_e }{dT_e d\Omega} = 
    \begin{cases}
    \dfrac{1.799\times 10^{44} T_e^{-12.061}}{1+2.762\times 10^{36} T_e^{-9.269}+3.853\times 10^{40} T_e^{-10.697}} &~{\rm if~ } T_e<6880 {\rm ~MeV}\\
    3.259 \times10^{10} T_e^{-3.505} +3.204\times10^5 T_e^{-2.620} &~{\rm if~ } T_e \geq 6880 {\rm ~MeV} 
    \end{cases}
\end{equation}
in units of $ {\rm sr}^{-1} {\rm m}^{-2} {\rm s}^{-1} {\rm MeV}^{-1}$, where $T_e$ is given in MeV.
Whereas for the CRp, we use the flux given in  \cite{DellaTorre:2016jjf,Boschini:2017fxq}. 
Apart from protons, there is an abundant number of helium nuclei in the CRs, which may also interact with DM if their nuclear interaction is allowed. Hence, in CRp BDM flux, two dominant contributions come from proton and helium nuclei ($^4$He).    
This leads us to explore specific details of the scattering cross-section between dark matter (DM) and cosmic rays (CR). In the rest frame of $\chi$, the energy-independent differential cross-section between CR particle $i$ and $\chi$, as commonly done in the literature, is given by~\cite{Bringmann:2018cvk}, 

\begin{equation}
    \frac{d\sigma_{\chi i}}{dT_\chi}=\frac{\sigma_{\chi i}}{T_\chi^{\rm max}},~~~{\rm where}~~T_\chi^{\rm max} =\dfrac{T_i^2+2 m_i T_i}{T_i+(m_i+M_\chi)^2/(2 M_\chi)}.
    \label{eq:cons}
\end{equation}
However, this relation is valid for constant amplitudes in non-relativistic limits and the results change in the presence of energy-dependent cross-sections as shown in \cite{Bardhan:2022bdg}. This discrepancy leads to a slightly stronger bound for energy-dependent cross-sections for heavy mediators and a weaker bound for light mediators \cite{Bardhan:2022bdg}. The energy-dependent differential cross-sections for heavy mediators are given by,
\begin{align}
    \left(\frac{d\sigma_{\chi i}}{dT_\chi}\right)_{\rm scal} &= \sigma_{\chi i} \left(\dfrac{M_\chi}{4\mu_{i \chi}^2}\right) \dfrac{(2M_\chi+T_\chi)(2 m_i^2+M_\chi T_\chi)}{s_{\rm CR}T_\chi^{\rm max}}\\ 
    \left(\frac{d\sigma_{\chi i}}{dT_\chi}\right)_{\rm vect} &= \sigma_{\chi i} \left(\dfrac{M_\chi}{2\mu_{i \chi}^2}\right)  \dfrac{ (2 M_\chi (m_i+T_i)^2 -T_\chi ((m_i+M_\chi)^2+2 M_\chi T_e)+M_\chi T_\chi^2 )}{s_{\rm CR} T_\chi^{\rm max} },
    \label{eq:deps}
\end{align}
where the first (second) one describes the heavy scalar (vector) mediated $\chi-i$ interaction. $s_{\rm CR}$ signifies the center of momentum energy of $\chi-i$ collisions, whereas $\mu_{i\chi}$ is the reduced mass of $\chi-i$ system.


\cref{eq:cons,eq:deps} straightforwardly applies for $\chi-e$ scattering with the obvious replacements.
In contrast to the $\chi-e$ scattering, one has to include the scattering form factor $G_i^2(q^2)$ for DM-nucleon interactions, where $q$ is the momentum transfer ($q^2=2 M_\chi T_\chi$). 
Thus for CRp boosted scenario, the DM flux is obtained from \cref{eq:dm_flux} multiplied with an additional factor of $G_i^2(q^2)$ using cross-sections computed in point-like limit \cite{Bringmann:2018cvk}.
For simplicity, we consider the dipole form factor,
\begin{equation}
    G_i(q^2)=\frac{1}{(1+q^2/\Lambda_i^2)^2},
\end{equation}
where, we take $\Lambda_p=770$ MeV ($\Lambda_{\rm He}=410$ MeV) for proton ($^4$He) \cite{Bringmann:2018cvk,Angeli:2004kvy,Maity:2022exk}.
We further simplify the scenario assuming that, for spin-independent scattering, DM couples with protons and neutrons with similar strength ($\sigma_{\chi p}=\sigma_{\chi n}$).
Hence, the total cross-section can be taken as $\sim A^2\sigma_{\chi n} (\mu_{\chi n}/\mu_{\chi\mathcal{N} })^2$ for $\chi$-He scattering, where $\mu_{\chi\mathcal{N} }$ is DM-nucleus reduced mass.
Throughout this work, we assume that DM couples to either electrons or nucleons. Analysis of the situation where both cross-sections are relevant will be reported in our future work.

\subsection{Diffuse Supernova neutrino background scattering} 
So far we have discussed DM having nuclear or electronic interactions.
On the other hand, if one allows DM to couple with neutrinos ($\nu$), the DSNB constituting an abundant neutrino flux with energy $\lesssim 100$ MeV can significantly enhance the DM kinetic energy \cite{Ghosh:2021vkt,Majumdar:2021vdw}. 
Thus, while computing the BDM flux, the contributions from neutrino interactions should also be explored.
The theoretical prediction for the DSNB flux from the core-collapse supernova events for each flavour $\alpha (\alpha\equiv e,\mu,\tau)$ of $\nu$ can be obtained as \cite{DeGouvea:2020ang,Beacom:2010kk},
\begin{equation}
    \frac{d\Phi^\alpha_{\nu_\alpha}}{dE_\nu}=\int_0^{z_{\rm max}} dz \frac{R_{\rm CCSN}}{H(z)} F_{\nu_\alpha}(E_{\nu_\alpha})|_{E_\nu=(1+z)E_\nu^s},
\end{equation}
where $R_{\rm CCSN}$ is the rate of core collapse supernova (CCSN) events dependent on the star formation rate.
$R_{\rm CCSN}$ can be obtained by the empirical formula from ref.\cite{Horiuchi:2008jz}, assuming stars are in equilibrium i.e. their birth and death rates are equal. 
The expansion rate of the Universe is denoted by the Hubble parameter, $H(z)=H_0\sqrt{\Omega_\Lambda+\Omega_m(1+z)^3}$, where $\Omega_\Lambda=0.68,~\Omega_m=0.3$ and $H_0=67.45~{\rm Km~s}^{-1}{\rm Mpc}^{-1}$ \cite{Planck:2018vyg}. 
$F_{\nu_\alpha}(T_{\nu_\alpha})$ signify the energy spectrum of the emitted neutrino of flavour $\alpha$ from a single CCSN. Assuming a Fermi Dirac distribution, it is given by{\footnote{The $E$ independent prefactors come from normalising the total energy emitted.}} \cite{DeGouvea:2020ang,Beacom:2010kk},
\begin{equation}
 F_{\nu_\alpha}(E_{\nu_\alpha})= \frac{E_\nu^{\rm tot}}{6}   \frac{120}{7 \pi^4} \frac{E_\nu^2}{T_\nu^4} \dfrac{1}{e^{E_\nu/T_{\nu_\alpha}}+1},
 \label{eq:dist}
\end{equation}
where total neutrino energy emitted from a supernova is depicted by $E_\nu^{\rm tot}\approx 10^{53}$ erg and
$T_{\nu_\alpha}$ is the temperature (not to be confused with kinetic energy) of the neutrinos of each flavour (electron neutrino ($\nu_e$), muon neutrino ($\nu_\mu$) and tau neutrino ($\nu_\tau$) ) \cite{Beacom:2010kk}.
Since DSNB has not yet been detected experimentally, these parameters can be fixed only from numerical simulations of Supernova \cite{Horiuchi:2017qja}.
Nevertheless, we use the most commonly used values $T_{\nu_e}=6.6$ MeV, $T_{\bar{\nu}_e}=7$ MeV and $T_{\nu_{\mu/\tau}}=10$ MeV (and also their antiparticles)\footnote{The resulting flux with these parameters are allowed from the current bounds set by Super-K \cite{Super-Kamiokande:2013ufi}. in accordance with existing literature \cite{Das:2021lcr,DeRomeri:2023ytt,Das:2024ghw,DeGouvea:2020ang}.}
$E_\nu$ is the energy of the neutrino.
Finally, to account for the total number of CCSN events throughout cosmic history, we integrate it over red-shift $z$ from the present ($z=0$) to the maximum red-shift for star formation, $z_{\rm max}\sim 6$ \cite{Ghosh:2021vkt}.
\begin{figure}[!tbh]
\centering
\includegraphics[scale=0.3]{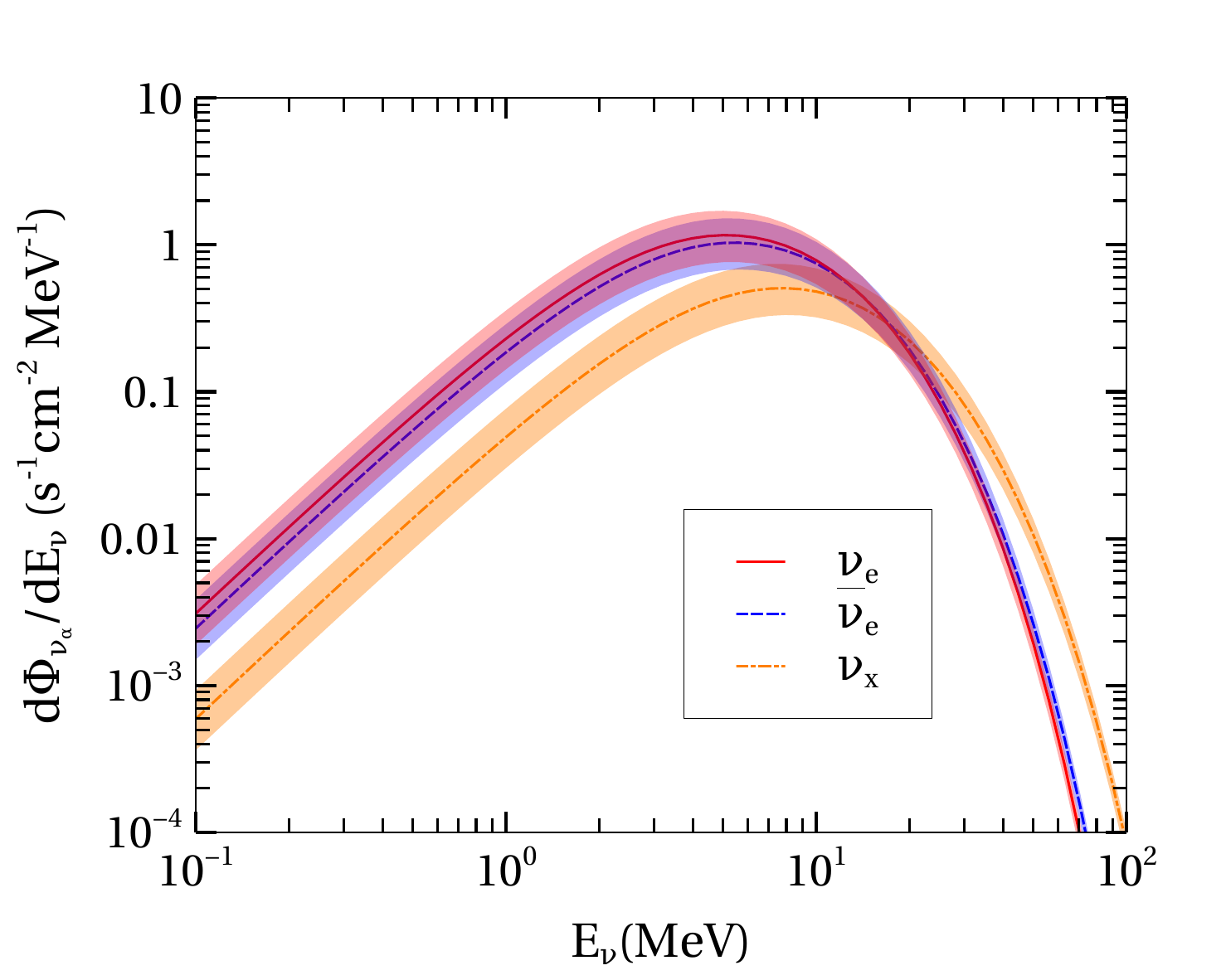}
\caption{Predicted differential flux of DSNB with respect to $E_\nu$ for each flavour of neutrinos \cite{DeRomeri:2023ytt}. The red solid, blue dashed, and orange dashed-dot lines signify the differential flux due to $\nu_e$, $\bar{\nu_e}$ and both $\nu_{\mu/\tau}$(and their antiparticle).
$\nu_{\mu/\tau}$ and $~\bar{\nu}_{\mu/\tau}$ are depicted by $\nu_x$.}
\label{fig:flux_DSNB}
\end{figure}
$E_\nu$ is the red-shifted neutrino energy, while $E_\nu^s$ is the neutrino energy at the source.
The main uncertainty in the predicted rate $R_{\rm CCSN}$ is indeed reflected in the DSNB flux with overall $40\%$ uncertainty, which should be taken care of while estimating the exclusion bounds \cite{DeRomeri:2023ytt}.
The predicted DSNB differential flux with neutrino energy $E_\nu$ is shown in \cref{fig:flux_DSNB}.
The red solid, blue dashed, and orange dashed-dot lines signify the differential flux contributions from $\nu_e$, $\bar{\nu_e}$, and $\nu_{\mu/\tau}$ (and $\bar{\nu}_{\mu/\tau}$) respectively.
Since $\nu_{\mu/\tau}$ (and their anti-particles) have been reported to have the same temperature by Super Kamiokande \cite{Super-Kamiokande:2013ufi}, we treat them identically denoted by $\nu_x$.
The behavior of the differential flux for all three flavours is mainly governed by the distribution function in \cref{eq:dist}. In the low-energy limit, $E_\nu\ll T_\nu$, the flux grows quadratically with $E_\nu$; whereas, in high-energy, it goes as $\exp(-E_\nu/T_\nu)$, leading to a peak like a pattern at around $E_\nu\sim T_\nu$. The lines are drawn assuming the mean value of the normalization in $R_{\rm CCSN}$ \cite{Beacom:2010kk,DeGouvea:2020ang}. 
However, the respective bands represent the uncertainties in the differential flux for each flavour after including the uncertainty in $R_{\rm CCSN}$.

The DSNB-boosted DM flux is obtained from the same \cref{eq:dm_flux} with the obvious replacements: $d\Phi_i^{\rm LIS}/dT_i \to d\Phi_\nu^{\rm tot}/dE_\nu $, $T_i\to E_\nu$, where $\Phi_\nu^{tot}$ denotes the total DSNB flux of all three neutrino flavours and their antiparticles.
The corresponding cross-section in \cref{eq:dm_flux} is replaced as,
\begin{equation}
    \frac{d\sigma_{\chi \nu}}{dT_\chi}=\frac{\sigma_{\chi \nu}}{T_\chi^{\rm max}(E_\nu)} \Theta(T_\chi^{\rm max}(E_\nu)-T_\chi),
\end{equation}
where, $T_\chi^{\rm max}= E_\nu^2/(E_\nu+M_\chi/2)$ \cite{Ghosh:2021vkt,DeRomeri:2023ytt}.

We now return to our previous discussion of BDM flux.
As hinted earlier, we assume that DM can interact either with both electron and neutrino ($\sigma_{\chi N}=0$) or with both nucleon and neutrino ($\sigma_{\chi e}=0$).
We identify the first scenario as electrophilic DM and the second scenario as nucleophilic DM.
We consider both CR and DSNB effects in calculating total BDM flux, i.e.
\begin{equation}
    \frac{d\Phi_\chi^{\rm tot}}{dT_\chi}=\left(  \frac{d\Phi_\chi}{dT_\chi}\right)_{CR} + \left(  \frac{d\Phi_\chi}{dT_\chi}\right)_{DSNB}.
\end{equation}

\begin{figure}[!tbh]
\centering
\subfigure[\label{a1}]{
\includegraphics[scale=0.32]{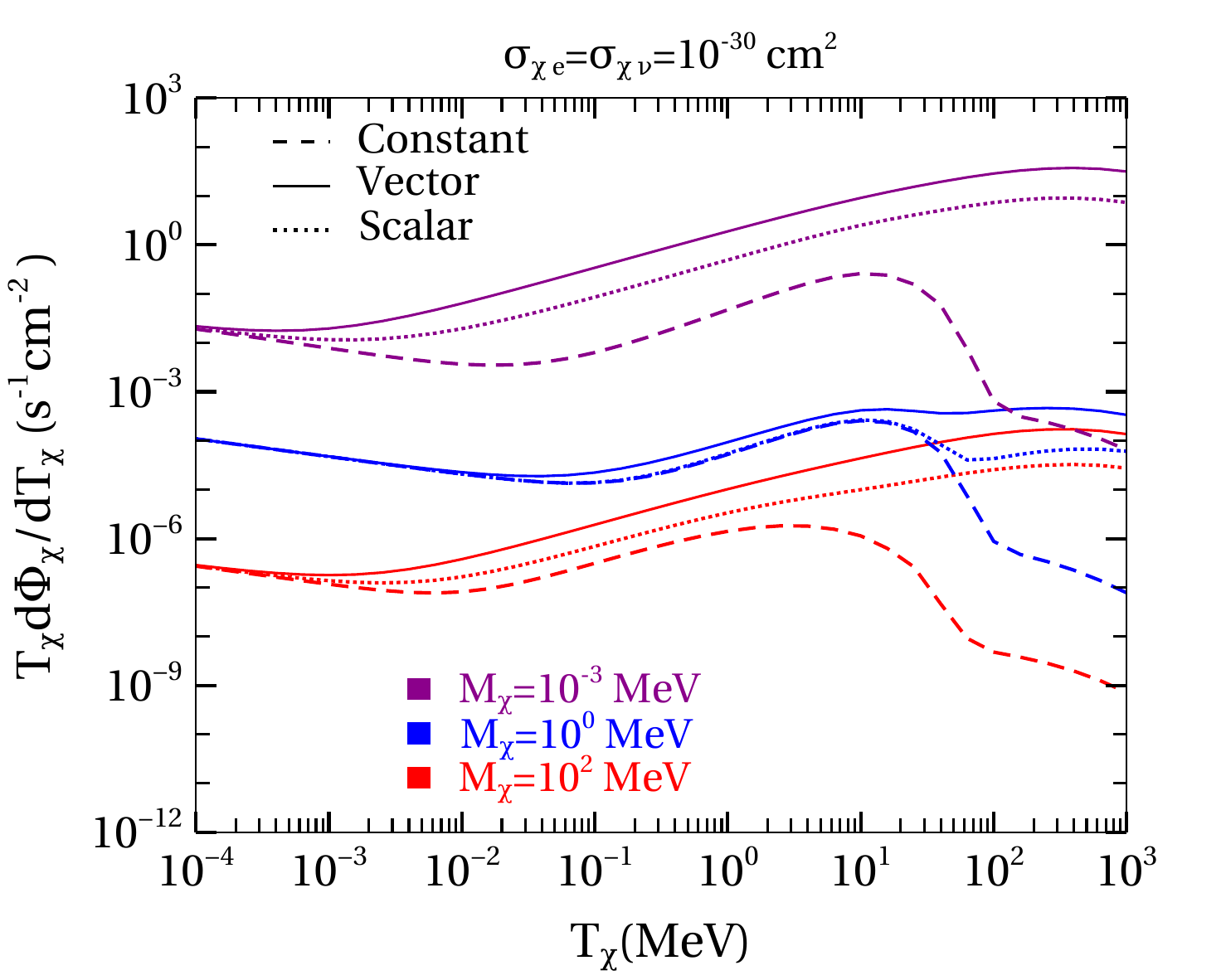}}
\subfigure[\label{a2}]{
\includegraphics[scale=0.32]{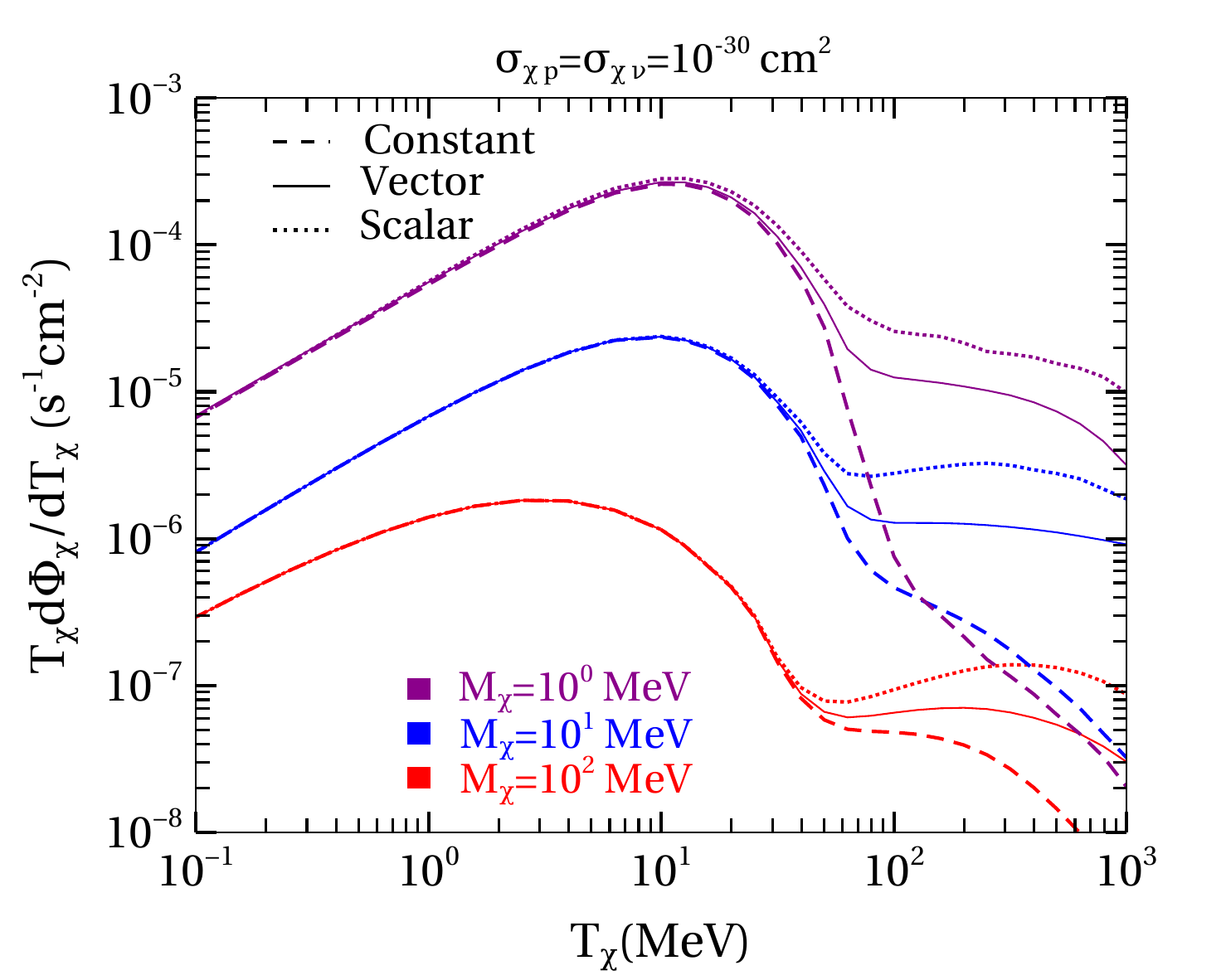}}
\caption{Boosted DM flux considering (a) cosmic ray electron and DSNB ; (b) cosmic ray nucleon and DSNB. We assume the same cross-section \cite{Ghosh:2021vkt} in both plots. }
\label{fig:flux_dsnb_CRe}
\end{figure}

In \cref{a1,a2}, we display the differential fluxes of boosted electrophilic and nucleophilic DM scenarios, respectively. \cref{a1} shows fluxes corresponding to DM masses $M_\chi=1$ keV (magenta), $M_\chi=1$ MeV (blue), and $M_\chi=100$ MeV (red) for a fixed value of cross-section, $\sigma_{\chi e} =\sigma_{\chi \nu}=10^{-30}~{\rm cm}^2$.
The dashed, dash-dotted, and solid lines indicate that the considered cross-sections are constant, scalar-mediated, and vector-mediated, respectively.
The DSNB fluxes obtain their peak value at $E_\nu\sim (1-10)$ (MeV) leading to a bump around $T_\chi \sim (0.1-10)$ (MeV) for a constant cross-section. 
In contrast, the CRe BDM fluxes for energy-dependent cross-sections (scalar and vector-mediated), the CRe BDM fluxes overshoot those with constant cross-section at $T_\chi\gtrsim 0.1$ (MeV) \cite{Bardhan:2022bdg}.
For this reason, CRe dominates over DSNB for energy-dependent cross-sections, especially in the low masses.

On the other hand, for nucleophilic DM in \cref{a2}, the magenta, blue, and red signify $M_{\chi}=1$ MeV, $M_{\chi}=10$ MeV, and $M_{\chi}=100$ MeV. 
We use the same convention to indicate the fluxes with energy-dependent and -independent cross-sections as shown in \cref{a1}. 
However, as the cosmic proton and He are less abundant than the predicted DSNB flux \cite{Bringmann:2018cvk}, we observe that the shape of differential flux is mostly dominated by DSNB flux.
As DSNB has maximum flux at $E_\nu$ around $\sim 10$ MeV, we see that up to $T_\chi\sim 10$ MeV, the lines with all types of cross-section overlap for a fixed $M_\chi$.
Since, DSNB flux is suppressed at $T_\nu\gtrsim10^2$ MeV, at $T_\chi\gtrsim10^2$ MeV, the CRp flux dominates.
Also, the differential flux starts to differ depending on the type of cross-sections in that particular region. 

After calculating the incoming BDM flux from different sources like CRe, CRp, and DSNB, we are now set to investigate the potential recoil rate of such DM particles in ground-based detectors. 
In the next section, we discuss the recoil in dedicated experiments due to BDM for both electrophilic and nucleophilic cases.

\section{BDM scattering in detectors}
\label{sec:recoil}
After the BDM reaches the ground-based detectors, it will scatter with electrons or nucleons depending on the scenario considered.
The differential recoil rate is given by,
\begin{equation}
    \frac{dR}{dE_R}=t_{\rm exp} N_{\rm T} \mathcal{E}(E_R) \int^{\infty}_{T_{\chi}^{\rm min}(E_R)} \frac{d\Phi_\chi^{\rm tot}}{dT_\chi} \frac{d \sigma_{\chi i}}{dE_R}dT_{\chi},
    \label{eq:recoil}
\end{equation}
where $E_R$ is the recoil energy, $t_{\rm exp}$
is the time of exposure and $N_T$ denotes the number of 
target particles. $T_{\chi}^{\rm min}$ is the minimum kinetic energy 
of $\chi$ required to generate a recoil energy $E_R$ and is given by
\cite{Bringmann:2018cvk,Bardhan:2022bdg},
\begin{equation}
    T_{\chi}^{\rm min}(E_R) =
    \left(\frac{E_R}{2}-M_\chi \right) \left[1 \pm \sqrt{1+\frac{2 E_R (m_i+M_\chi)^2}{m_i(2 M_\chi-E_R)^2}} \right] ,
    \label{eq:tmin2}
\end{equation}
where $(+)$ or $(-)$ sign is applicable when $E_R > 2M_\chi$ or $E_R < 2 M_\chi$, respectively.
$\mathcal{E}(E_R)$ is the efficiency factor of the experiment and $m_i$ stands for the mass of the nucleus or electron depending on the scenario.
The differential scattering cross-section in the target is given by,
\begin{align}
    \left(\frac{d\sigma_{\chi i}}{dE_R}\right)_{\rm cons.}&=\frac{\sigma_{\chi i}}{E_R^{\rm max}(T_\chi)},\\
    \left(\frac{d\sigma_{\chi i}}{dE_R}\right)_{\rm scal.} &= \sigma_{\chi i} \left (\dfrac{m_i}{4\mu_{i \chi}^2} \right ) \dfrac{(2m_i+E_R)(2 M_\chi^2+m_i E_R)}{s_{\chi}E_R^{\rm max}},\\ 
    \left(\frac{d\sigma_{\chi i}}{dE_R}\right)_{\rm vect.} &= \sigma_{\chi i} \left (\dfrac{m_i}{2\mu_{i \chi}^2}\right )\dfrac{ \left [2 m_i (M_\chi+E_R)^2 -E_R \left ((m_i+M_\chi)^2+2 m_i T_\chi\right )+m_i E_R^2 \right ]}{s_{\chi}E_R^{\rm max} },
\end{align}
for constant cross-section, scalar-mediated cross-section, and vector-mediated cross-section, respectively.
Here $s_\chi$ is the center of mass energy squared of the DM--target particle system and $E_R^{\rm max}$ is the maximum recoil energy possible. 
They are given by
\begin{equation}
    E_R^{\rm max} =\dfrac{T_\chi^2+2 M_\chi T_\chi}{T_\chi+(M_\chi+m_i)^2/(2 m_i)},~~~s_{\chi}=(M_\chi+m_i)^2+2 m_i T_\chi.
\end{equation}
For electronic recoil in XENONnT \cite{XENON:2023cxc} to account for the binding energies of the bound electron, $N_T$ is given by,
\begin{equation}
    N_T= N_A \sum_{i=1}^Z \Theta(E_R-E_{B_i}),
\end{equation}
where $Z$ and $N_A$ signify the number of electrons in an atom and the total number of atoms in the detector, respectively.
$E_{B_i}$ stands for the binding energy of the $i^\text{th}$ electron in xenon atom.
The Heaviside function dictates the number of electrons that can be ionized with recoil energy, $E_R$.
The binding energies of the single bound electrons are provided in \cite{Chen:2016eab}, which uses the Hartree-Fock method. 
With these $N_T$, we consider the electron as a free particle to evaluate the scattering cross-section \cite{Chen:2016eab,Majumdar:2021vdw,DeRomeri:2023ytt}.
Since the threshold energy for Super-Kamiokande is well above the binding energies, there we can simply treat electrons as free particles with $N_T= Z \times N_A$ \cite{Super-Kamiokande:2011lwo}. 

However, computing the nuclear recoil rate (for nucleophilic DM) is slightly less straightforward.
To compare with the observed data, it is customary to express the nuclear recoil rate in terms of electron equivalent energy \cite{DeRomeri:2023ytt,Maity:2022exk}. 
The nuclear recoil energy ($E_N$) is related to equivalent electron recoil energy ($E_R$) as \cite{Essig:2018tss}
\begin{equation}
    E_R= \mathcal{Q}(E_N) E_N,
\end{equation}
where $Q(E_R^N)$ is the nuclear quenching factor. For this work, we take the Lindhard quenching factor given by \cite{osti:4701226}
\begin{equation}
    Q(E_N)=\dfrac{k(1+3\epsilon^{0.15}+0.7\epsilon^{0.6} +\epsilon)}{1+k(1+3\epsilon^{0.15}+0.7\epsilon^{0.6} +\epsilon)},
\end{equation}
where $\epsilon= 11.5 Z^{-7/3} (E_N/{\rm keV})$ ($Z$ atomic number) and the value of $k$ used in literature is $k=0.145$ \cite{Maity:2022exk}.
Using the formula of partial derivatives, the differential recoil rate with respect to the electron equivalent energy is given by \cite{Maity:2022exk,DeRomeri:2023ytt},
\begin{equation}
    \frac{dR}{dE_R}= \frac{dR}{d E_N} \frac{1}{Q(E_N)+ E_N \frac{dQ}{dE_N}}.
\end{equation}
Note that to evaluate $dR/dE_N$, \cref{eq:recoil} can be used with the obvious replacement $E_R\to E_N$, and the efficiency factor is also considered as a function of $E_N$ \cite{XENON:2023cxc}. $N_T$ is given by simply the number of nucleons in the detector.

\begin{figure}[!tbh]
    \centering
    \subfigure[\label{s1}]{
    \includegraphics[scale=0.29]{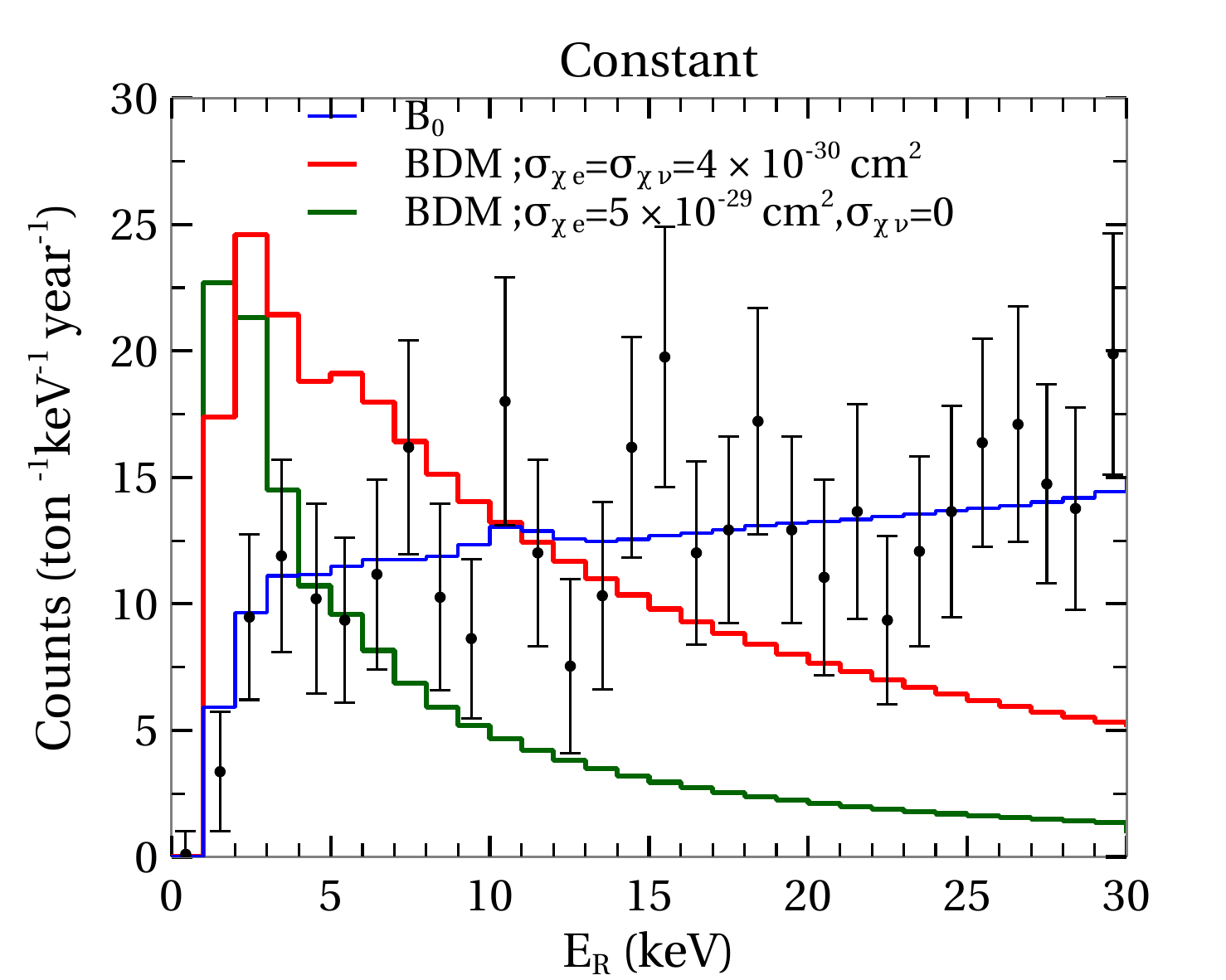}}
    \subfigure[\label{s2}]{
    \includegraphics[scale=0.29]{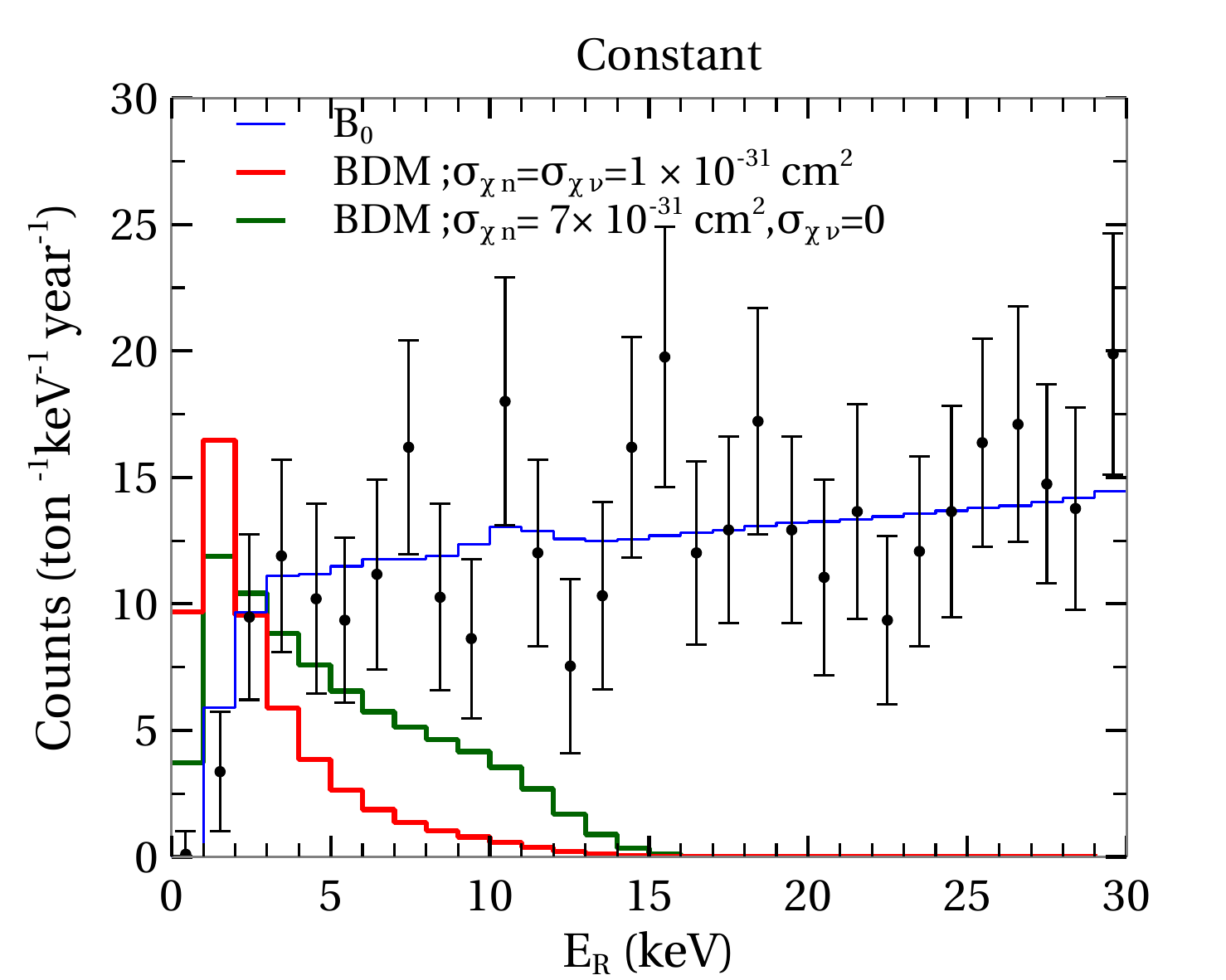}}
    \subfigure[\label{s3}]{
    \includegraphics[scale=0.29]{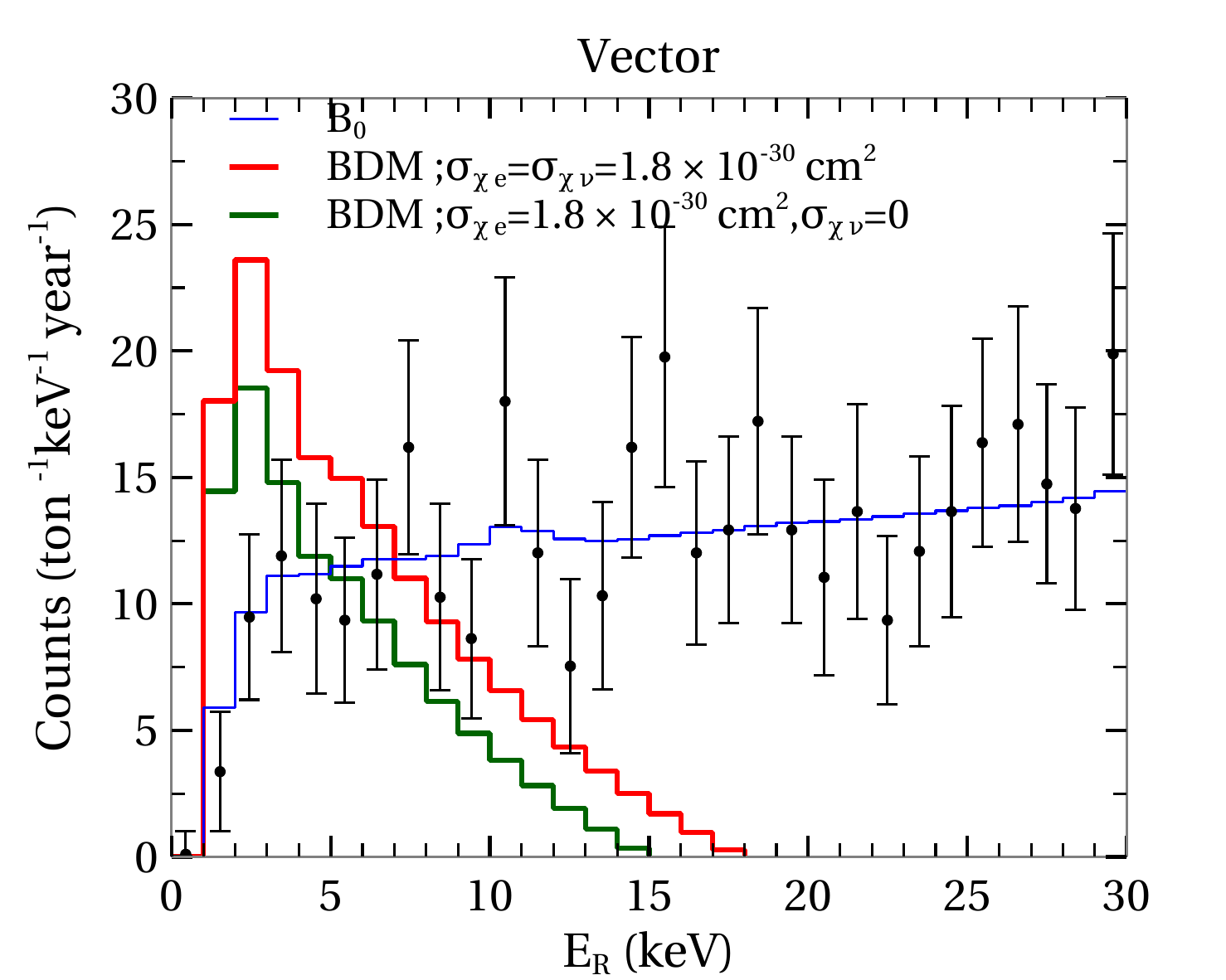}}
    \subfigure[\label{s4}]{
    \includegraphics[scale=0.29]{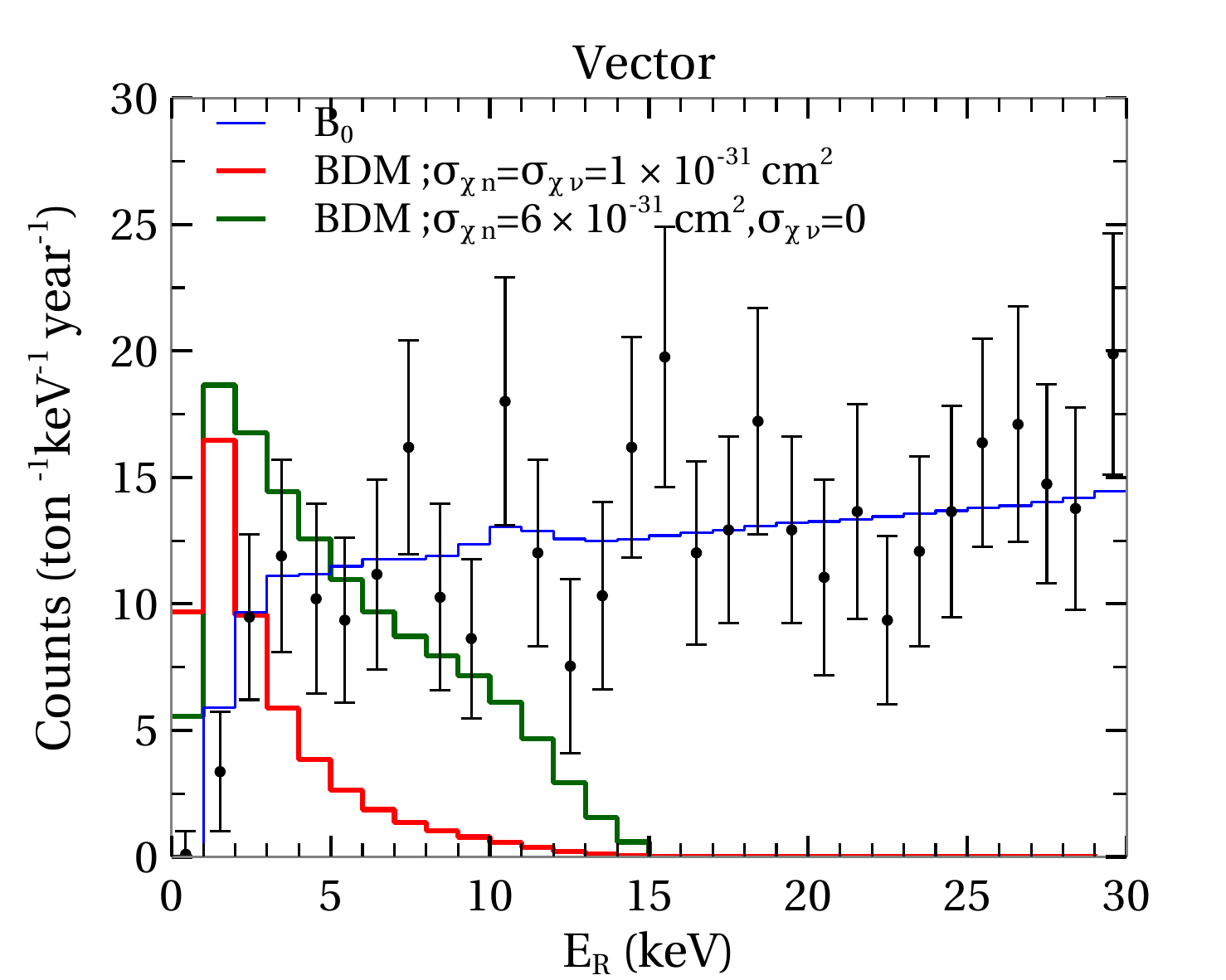}}
     \subfigure[\label{s5}]{
    \includegraphics[scale=0.29]{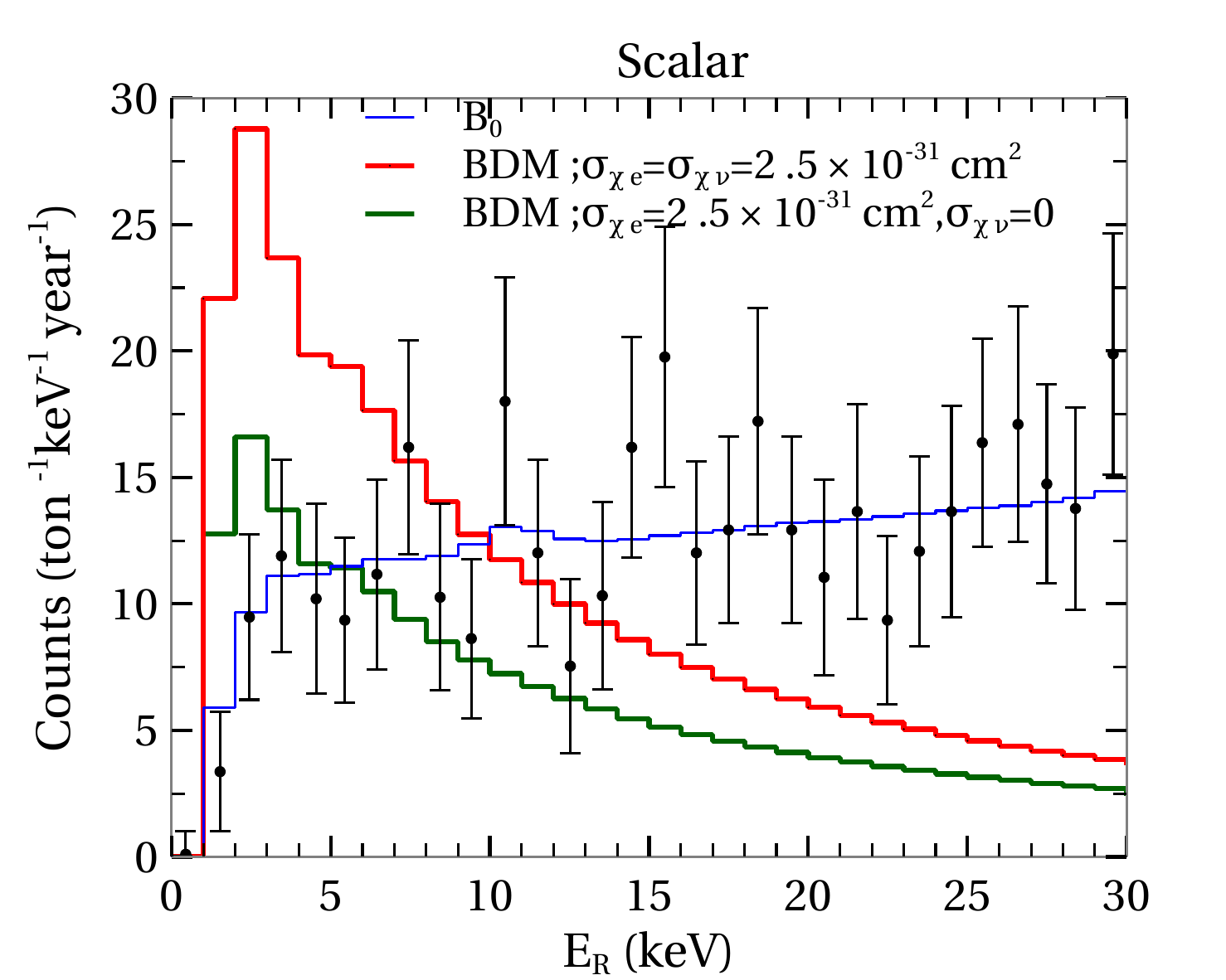}}
    \subfigure[\label{s6}]{
    \includegraphics[scale=0.29]{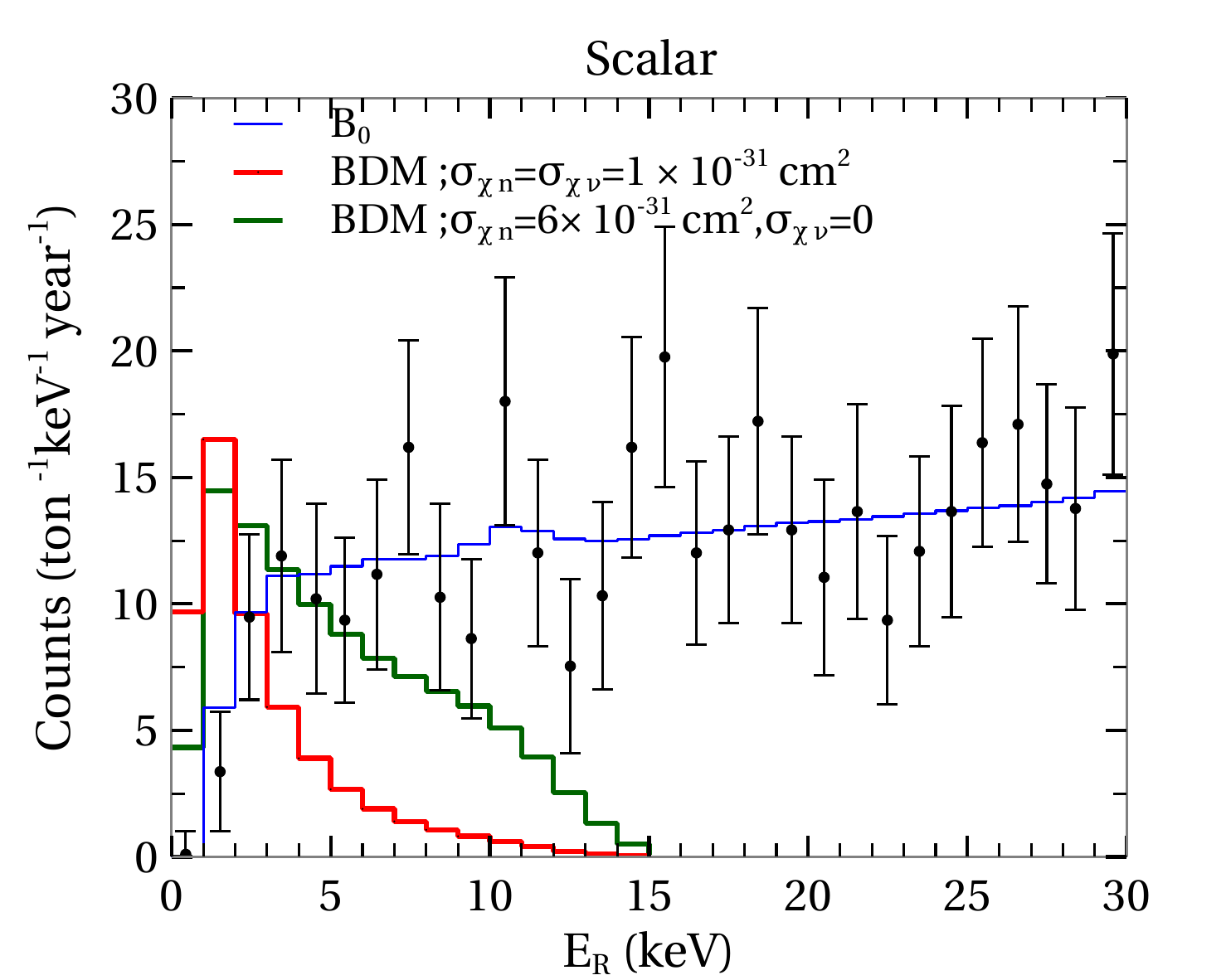}}
    \caption{Expected recoil rate in XENONnT experiment with year long exposure. {\bf Left} panels show the signal for electrophilic DM, whereas {\bf right} panels show the case for nucleophilic DM. The Blue curve is the expected background event rate and the black dots with $1\sigma $ error bars are the observed event rate from \cite{XENON:2022ltv}.}
    \label{fig:rec_rate}
\end{figure}

In \cref{fig:rec_rate}, we show the differential event rate as a function of electron equivalent recoil energy, $E_R$ for electrophilic and nucleophilic DM for $M_{DM}=100$ MeV. The plots in the left and right columns contain results for electrophilic ($\sigma_{\chi n}=0$)  and nucleophilic ($\sigma_{\chi e}=0$) DM, respectively.
For all plots, black dots and blue lines indicate the observed data  with $1\sigma$ error bars  and background, respectively.
Moreover, the red solid lines signify the event rate for DM boosted by both CR and DSNB (benchmark: $\sigma_{\chi \nu}=\sigma_{\chi e}$ for electrophilic and $\sigma_{\chi \nu}=\sigma_{\chi n}$ for nucleophilic DM). 
On the other hand, green lines stand for the event rate generated by DM boosted by CR only ($\sigma_{\chi \nu}=0$).
As mentioned earlier both CRe boosted DM flux as well as constraints differ significantly while considering energy-dependent interactions (scalar and vector-mediated) compared to constant cross-section \cite{Bardhan:2022bdg}.
Motivated by this, we also consider different interactions like constant, vector, and scalar cross-sections.
The results for DM  with constant cross-sections, heavy vector-mediated interactions, and heavy scalar-mediated interactions are shown in the top, middle, and bottom rows, respectively.
Note that, with an increase in incoming DM energy $T_\chi$, the DM flux decreases as shown in \cref{fig:flux_dsnb_CRe}. 
On the other hand, producing higher recoil energy, $E_R$ requires high incoming DM energy, $T_\chi$ as discussed earlier in this section.
For this reason, with an increase in $E_R$, the event rate decreases for all the plots in \cref{fig:rec_rate} as the event rate is proportional to the incoming flux (see \cref{eq:recoil}).
However, apart from the incoming flux another crucial input that governs the event rate in \cref{eq:recoil} is the detector efficiency factor $\mathcal{E}$.
For both electron and nucleon, the efficiency factor drops and tends to zero at $E_R\lesssim 0.5$ keV \cite{XENON:2022ltv,XENON:2023cxc}. The combined effect of incoming flux and efficiency leads to a peak-like behaviour in the event rate at low $E_R$ for all the plots in \cref{fig:rec_rate}.

We first discuss the plots in the left panel of \cref{fig:rec_rate}, the electrophilic DM scenario. The event rates increase significantly for all three types of interactions i.e. constant, vector, and scalar when the effect of DSNB is included as shown in \cref{s1,s3,s5}, respectively.
For example, in \cref{s1}, we plot the DM signal boosted by CRe only (green line) with the benchmark cross-section, $ \sigma_{\chi e }=5\times10^{-29}~{\rm cm}^2$ (and $\sigma_{\chi \nu}=0$), which corresponds to the existing constraint on constant $\sigma_{\chi e}$ for $M_\chi=100$ MeV, as analyzed in \cite{Bardhan:2022bdg}.
Despite having an order smaller cross-sections $(\sigma_{\chi \nu}=\sigma_{\chi e }=4\times10^{-30}~{\rm cm}^2)$ along with both DSNB and CRe, the BDM signal rate (red line) remains competitive with signals from CRe-boosted DM. Thus, we expect that DM interactions mediated by both DSNB and CRe, with similar interaction strengths, can access unexplored regions of parameter space.
We consider $\sigma_{\chi e}= 1.8 \times 10^{-30}~{\rm cm}^2$ for vector- (\cref{s3}) and $\sigma_{\chi e}= 2.5 \times 10^{-31}~{\rm cm}^2$ (\cref{s5}) for scalar-mediated interaction.
The event rates are substantially modified by the combined effects of DSNB and CRe boosts, for both vector- and scalar-mediated DM interactions, as seen in \cref{s3} and \cref{s5}, respectively.

Now we turn our attention to the right panel of \cref{fig:rec_rate}, the nucleophilic DM scenario.
Including the DSNB effect also leads to higher events for nucleophilic DM, as shown in \cref{s2,s4,s6}. We compare the quenched nuclear recoil spectrum to the observed spectrum in the XENONnT electron recoil analysis \cite{XENON:2022ltv}. The electron recoil search ignores events in the nuclear recoil band. However, the observed event rate in the nuclear recoil band \cite{XENON:2023cxc} is very low compared to the electron recoil search. Consequently, the limit we obtain with this method is conservative.
For DSNB and CRp BDM, we consider $ \sigma_{\chi \nu}=\sigma_{\chi e }=1\times10^{-31}~{\rm cm}^2$ for three plots in right panel.
For DM boosted by CRp only ($\sigma_{\chi \nu}=0$), we consider  $\sigma_{\chi e}= 7 \times 10^{-31}~{\rm cm}^2, ~6 \times 10^{-31}~{\rm cm}^2$ and $6 \times 10^{-30}~{\rm cm}^2$, respectively for constant (\cref{s2}), vector-mediated (\cref{s4}) and scalar-mediated (\cref{s6}) interactions, respectively.
Again, for all three cases even with smaller cross-sections the event rates due to both CRp and DSNB boosted DM are competitive to events with CRp effects only.
The enhanced rate after including the boost, by both CRp and DSNB, will lead to stronger constraints on cross-sections as we will see later.
However, for the nucleophilic case, for $M_{\rm DM}=100$ MeV the BDM flux is mostly dominated by the contribution from DSNB (see \cref{a2}). For this reason, we see no significant change in the event rates for energy-dependent cross-sections when $\sigma_{\chi \nu}=\sigma_{\chi n}$ (red lines) in the right column. It is worth noting that while we use the mean value for the DSNB flux, the DSNB-boosted event rate may vary in the presence of uncertainty.
In our analysis, we account for this uncertainty in the predicted DSNB flux when generating the parameter space exclusion plots. 

Thus, including the DM-$\nu$ cross-section ($\sigma_{\chi \nu}\neq0$) may open the possibility of DM being scattered by DSNB apart from the CR particles which may lead to the enhanced event rate.
Also, with energy-dependent cross-sections, the DM signals change significantly for electrophilic DM and are indifferent to nucleophilic DM.
Keeping this in mind, in the next section, we display the constraints on the parameter space of DM boosted by CR and DSNB.

\section{Results}
\label{sec:result}
In this section, we perform a $\chi^2$ analysis and showcase the exclusion region in DM parameter space from the two ground based detectors: XENONnT and Super-Kamiokande (Super-K).
To carry out the $\chi^2$ analysis for XENONnT, we use the following expression for $\chi^2$ as defined in \cite{Almeida:1999ie,DeRomeri:2023ytt},
\begin{equation}
    \chi^2= \sum_i \left(\dfrac{R^i_{\rm pred}(\beta) - R^i_{\rm exp}}{\sigma_i}\right)^2 +\left(\frac{\beta}{\sigma_\beta}\right)^2,
\end{equation}
where, $R^i_{\rm pred}= (1+\beta) R^i_{\rm BDM}+B_0^i$ with  $R^i_{\rm BDM}$ and $B_0^i$ signifying the predicted recoil rate and simulated background, respectively in the $i^{\rm th}$ energy bin. 
Meanwhile, $R^i_{\rm exp}$ denotes the experimentally observed event rate, and $\sigma_i$ is the associated uncertainty in the observed data in $i^{\rm th}$ energy bin.
The parameter $\beta$ is the nuisance parameter to account for the associated uncertainty in the DSNB flux given by $\sigma_\beta=40\%$. For each new physics parameter sets, denoted by $\{\sigma_{\chi i}, M_{\rm DM}\}$, $\chi^2$ values are marginalized.
For each set of parameter points, the nuisance parameter $\beta$ is varied to obtain the minimum values of $\chi^2$ for each set $\{\sigma_{\chi i}, M_{\rm DM}\}$. Now, from the list of $\{\sigma_{\chi i}, M_{\rm DM}, \chi^2\} $ the minimum value of $\chi^2$ is referred as $\chi^2_{min}$ and $\Delta \chi^2=(\chi^2-\chi^2_{min})$ is obtained corresponding to each set of parameter points. The $90\%$ confidence level (C.L.) exclusion region is found by the criterion $\Delta \chi^2 > 2.71$ \cite{reid,Maity:2022exk}.
For XENONnT, the observed data and background are taken from ref.\cite{XENON:2022ltv}.

For Super-K, we follow a similar approach with experimental inputs taken from ref.\cite{Super-Kamiokande:2011lwo}. In contrast to XENONnT, Super-K has reported their observed event rate with a bin size of 4 MeV starting at $E_R=16$ MeV.
Since, for Super-K, the estimate of background $B_0^i$ is not found, we assume it to be equal to the observed rate $R_{\rm exp}^i$, as commonly done in literature \cite{Bardhan:2022bdg}.
Even though Super-K is sensitive at higher recoil energy ($\sim$ MeV) compared to XENONnT, with the large detector size, Super-K places stringent constraints on DM as will be shown in this section. 

\begin{figure}[!tbh]
\centering
\subfigure[\label{e1}]{
\includegraphics[scale=0.32]{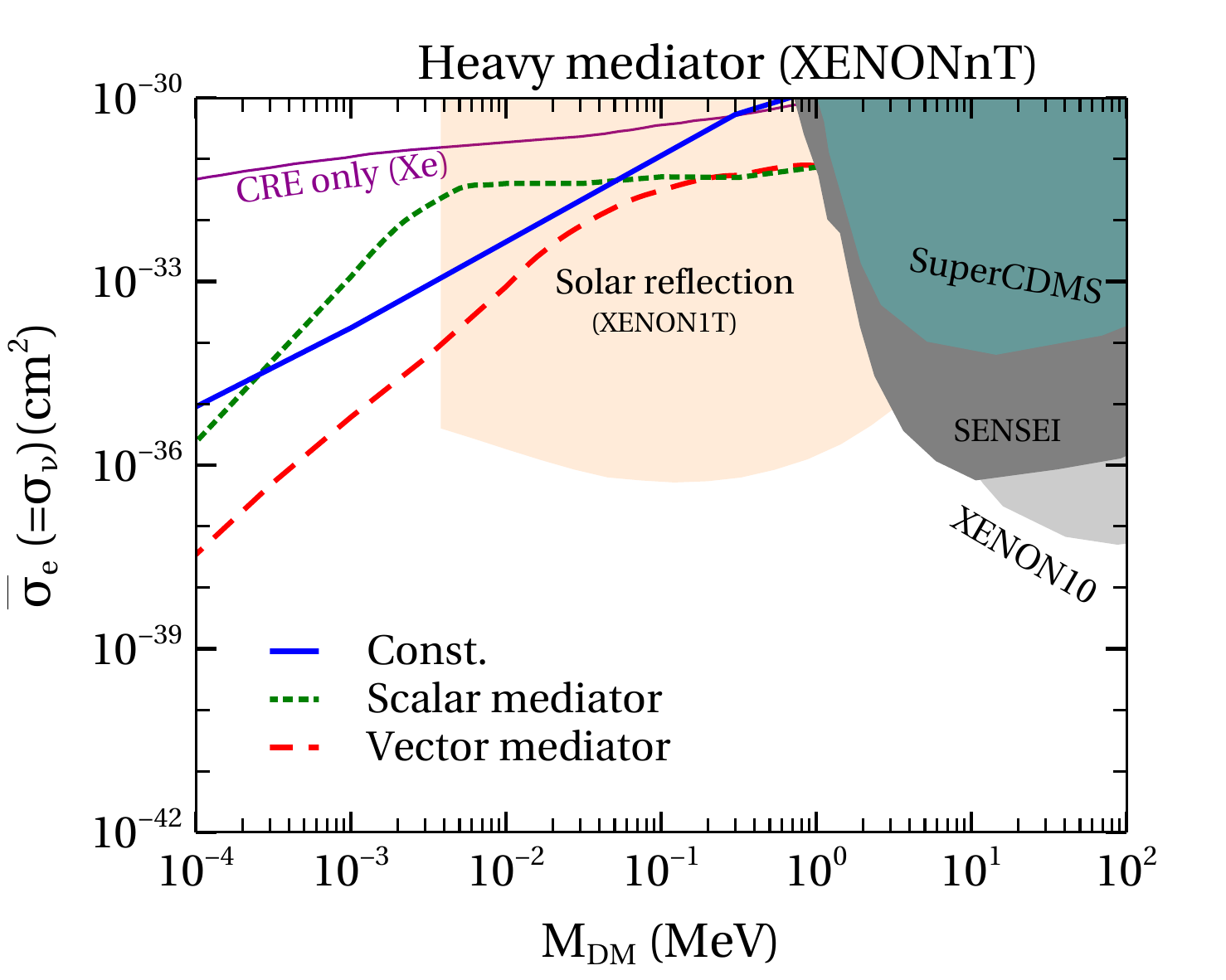}}
\subfigure[\label{e2}]{
\includegraphics[scale=0.32]{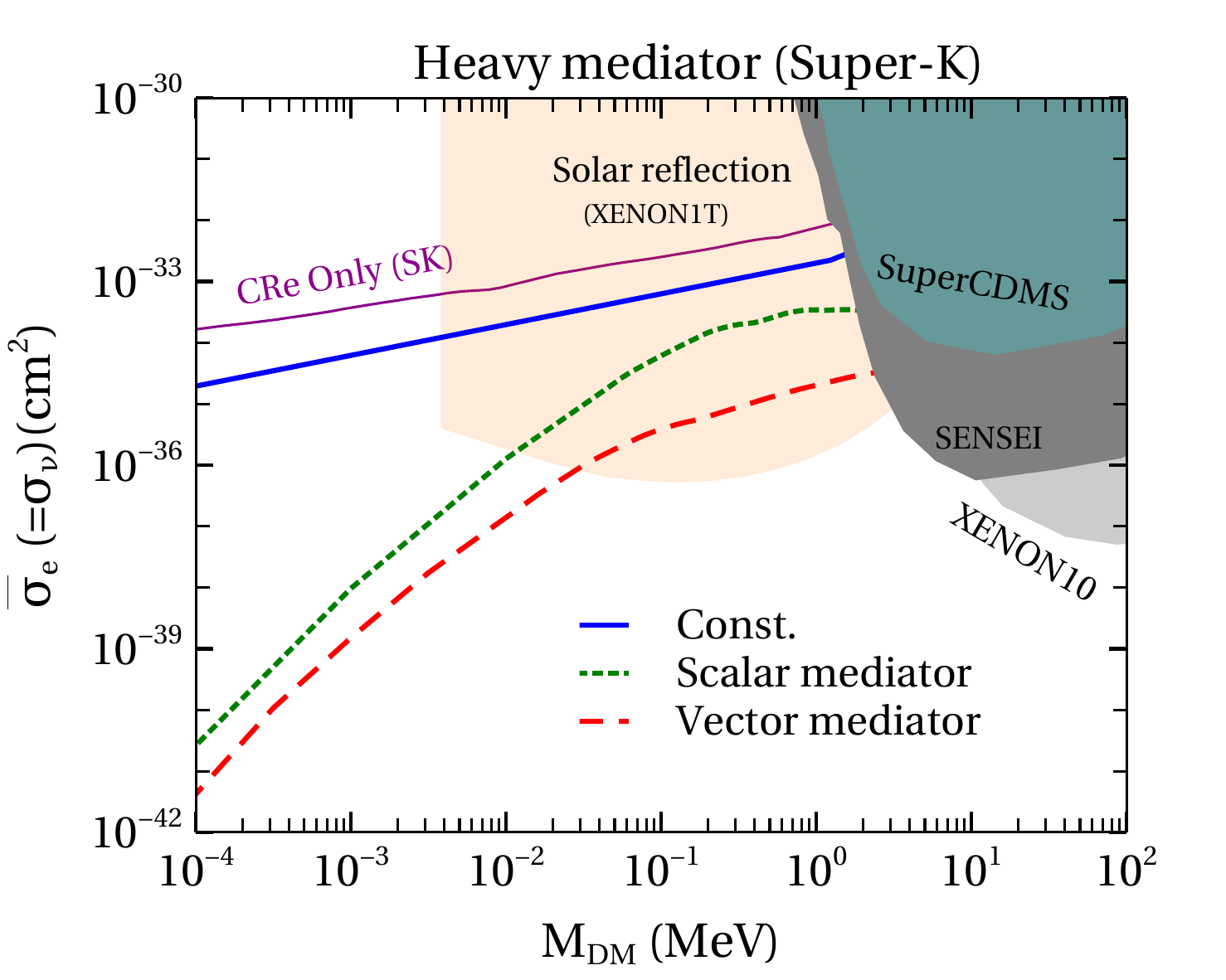}}
\caption{ Constraints on $M_{DM}$ vs. $\sigma_{\chi e}(=\sigma_{\chi \nu })$ plane at $90\%$ C.L from (a) XENONnT  (b) Super-Kamiokande assuming heavy mediator. Limits corresponding to constant cross-section as well as
scalar, and vector-mediated cross-sections are shown in blue, green, and red, respectively.}
\label{fig:elec_cons}
\end{figure}

In \cref{fig:elec_cons}, we show the constraints on $M_{DM}$ vs. $\sigma_{\chi e}$ plane for the electrophilic DM case.
We have taken the effects of both CRe and DSNB assuming $\sigma_{\chi e}=\sigma_{\chi \nu}$ as commonly used in existing literature \cite{Das:2021lcr,Ghosh:2021vkt,DeRomeri:2023ytt,Das:2024ghw}. 
The constraints obtained with our analysis from XENONnT and Super-K are shown in \cref{e1} ({\it left}) and \cref{e2} ({\it right}), respectively.
Blue, green, and red lines depict the constraints considering constant (energy-independent), scalar-mediated, and vector-mediated (energy-dependent) cross-sections.
The other existing constraints from SuperCDMS \cite{SuperCDMS:2018mne}, SENSEI \cite{Crisler:2018gci}, and XENON10 \cite{Essig:2012yx} are also portrayed for comparison.
Electron recoil experiments have a relatively lower threshold compared to nuclear recoil experiments, and hence, we observe that the existing constraints are sensitive up to 1 MeV.
Despite all the stringent constraints, our obtained bounds exclude some new parameter space and are tighter compared to the bounds obtained when only one of the two effects: CRe and DSNB is considered. 
The constraints are competitive to the bound obtained assuming BDM from solar reflection with the light orange shaded region \cite{An:2017ojc}.
Finally, we highlight that considering energy dependent cross-sections may lead to more sensitivity for heavy mediators as shown in \cref{fig:elec_cons} compared to the existing literature \cite{Ghosh:2021vkt}.
The situation is more promising for Super-K (\cref{e2}) where it can exclude a significant amount of new parameter space due to its larger detector volume.

\begin{figure}[!tbh]
\centering
\subfigure[\label{c1}]{
\includegraphics[scale=0.32]{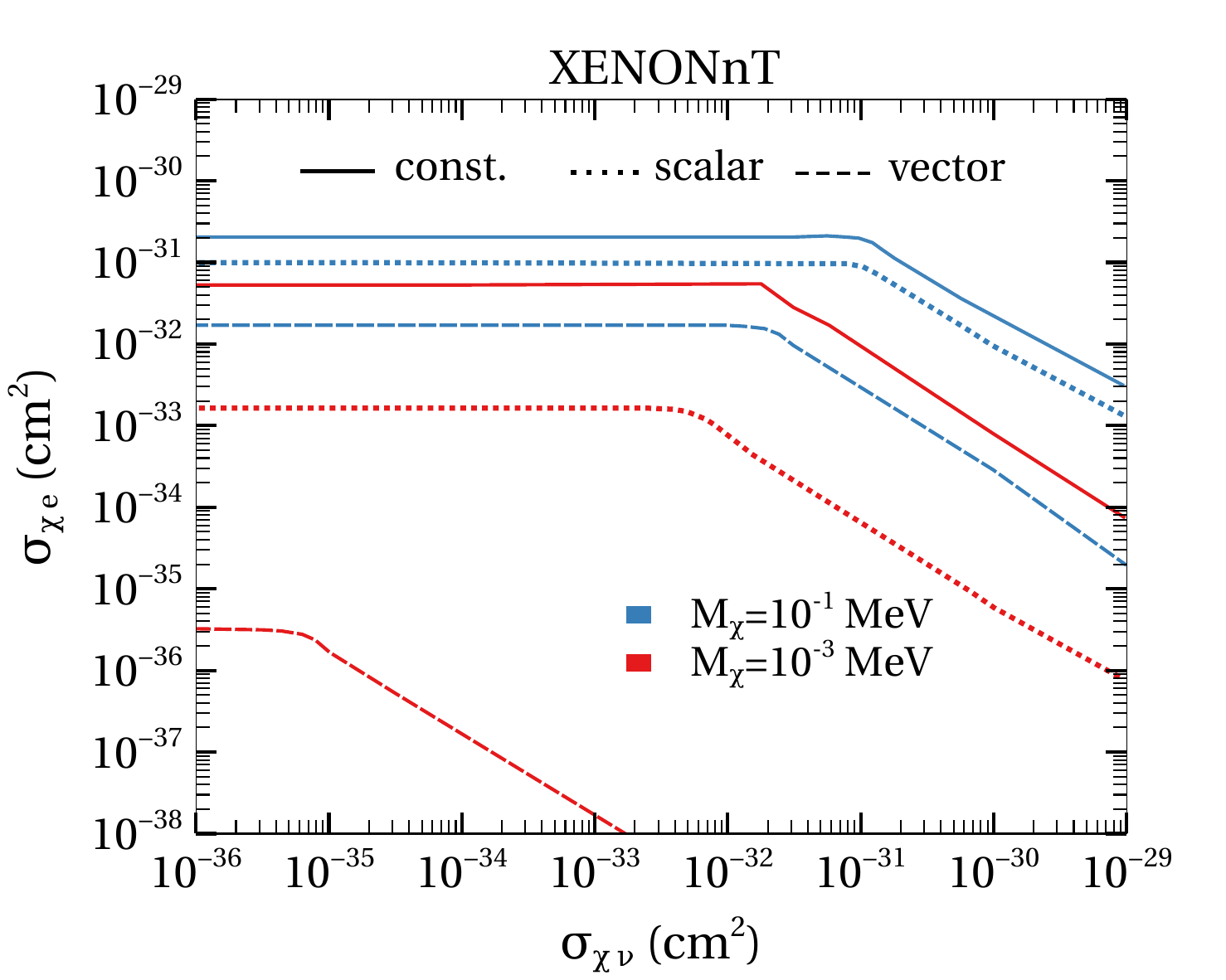}}
\subfigure[\label{c2}]{
\includegraphics[scale=0.32]{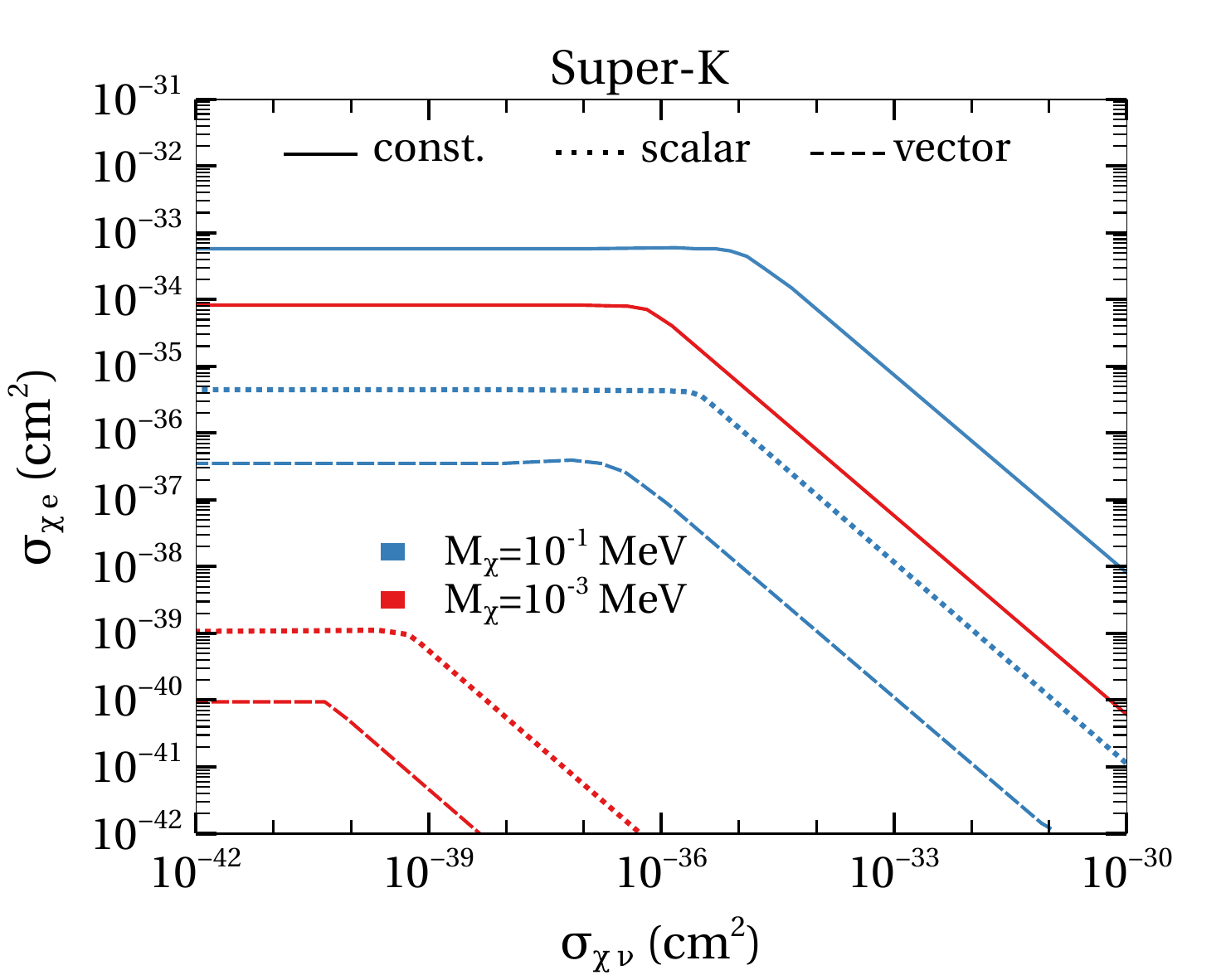}}
\caption{Contour lines of $90\%$ C.L. constraints on the $\sigma_{\chi \nu}$ vs. $\sigma_{\chi e}$ plane for different values of $M_\chi$ from (a) XENONnT and (b) Super-Kamiokande. The solid, dashed, and dotted lines represent constant, vector, and scalar cross-sections. The constraints corresponding to different $M_\chi\in \{10^{-3},10^{-1}\}$ MeV are shown by red and light blue colour, respectively. The regions above the contour lines are excluded.}
\label{fig:cont_e}
\end{figure}
In \cref{fig:cont_e}, we display the $90\%$ C.L. constraints on the $\sigma_{\chi \nu}$ vs. $\sigma_\chi e$ plane for individual $M_\chi$ from XENONnT (\cref{c1}) and Super Kamiokande (\cref{c2}).
The constraints corresponding to different interactions: constant, vector-, and scalar-mediated are depicted by solid, dashed, and dotted lines, respectively.
The different values of $M_\chi \in \{10^{-3},10^{-1}\}$ MeV are shown in red and light blue colour, respectively.
To roughly understand the pattern of the contour lines we refer to \cref{eq:recoil}, where we see 
 that the recoil rate is dependent on the cross-sections like,
\begin{equation}
    R\propto \Phi_{CRe} \sigma_{\chi e}^2 +\Phi_{DSNB} \sigma_{\chi e} \sigma_{\chi \nu}.
\label{eq:rate_sig}
\end{equation}
$\Phi_{CRe}$, and $\Phi_{DSNB}$ represents CRe and DSNB flux, respectively.
It should be emphasized that, since DSNB flux is higher than the CRe flux by only factor $\sim10$ (in the MeV energy \cite{Bardhan:2022bdg}), in the limit $\sigma_{\chi \nu}\ll \sigma_{\chi e}$, the first term in \cref{eq:rate_sig} dominates over the second term. For this reason, the contour becomes flat at a very small value of $\sigma_{\chi \nu}$. 
At this limit, the constraints on $\sigma_{\chi e}$ are solely due to CRe boosted DM and agree well with the existing literature \cite{Bardhan:2022bdg}.
However, when $\sigma_{\chi \nu}$ is high ($\gg \sigma_{\chi e} $), the second term dominates in \cref{eq:rate_sig} leading to a constraint on the product $\sigma_{\chi e} \sigma_{\chi \nu}$ shown by the tilted part of the curves.
We must highlight that even moderate values of $\sigma_{\chi \nu} (\sim 10 \times \sigma_{\chi e})$ help to obtain stronger constraints on $\sigma_{\chi e}$ compared to the existing constraints from CRe-boosted DM alone. This improvement in constraints is the main focus of our work.
With lower DM mass, the higher flux leads to stronger constraints on the cross-sections.
A key finding of our work is that energy-dependent cross-section leads to different constraints for the same $M_\chi$. The difference is more prominent for lower DM masses.
Again, Super-K gives the best sensitivity for DM-e interaction (due to its larger volume) as is also evident in the figure above.

\begin{figure}[!tbh]
\centering
\subfigure[\label{n1}]{
\includegraphics[scale=0.32]{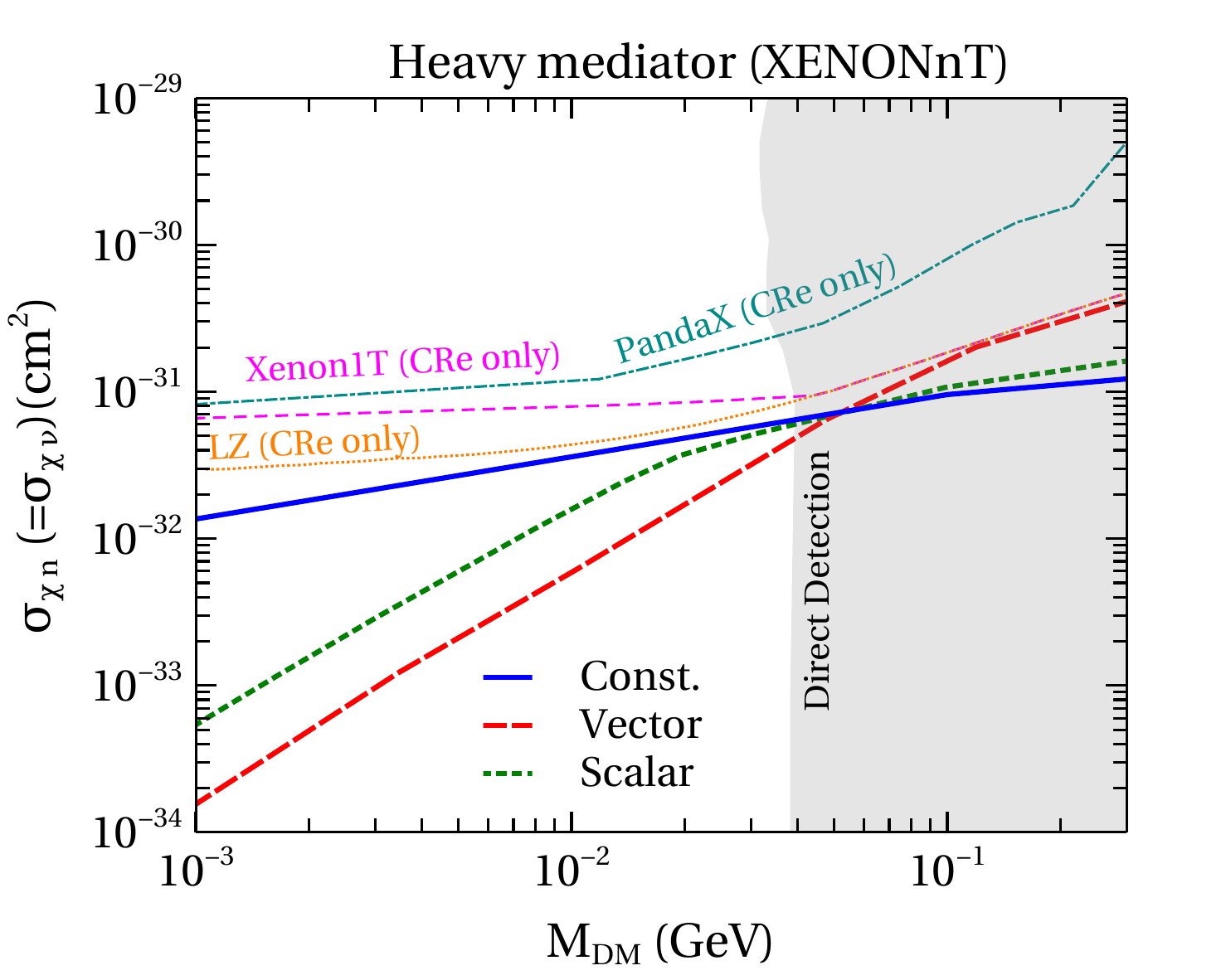}}
\subfigure[\label{n2}]{
\includegraphics[scale=0.32]{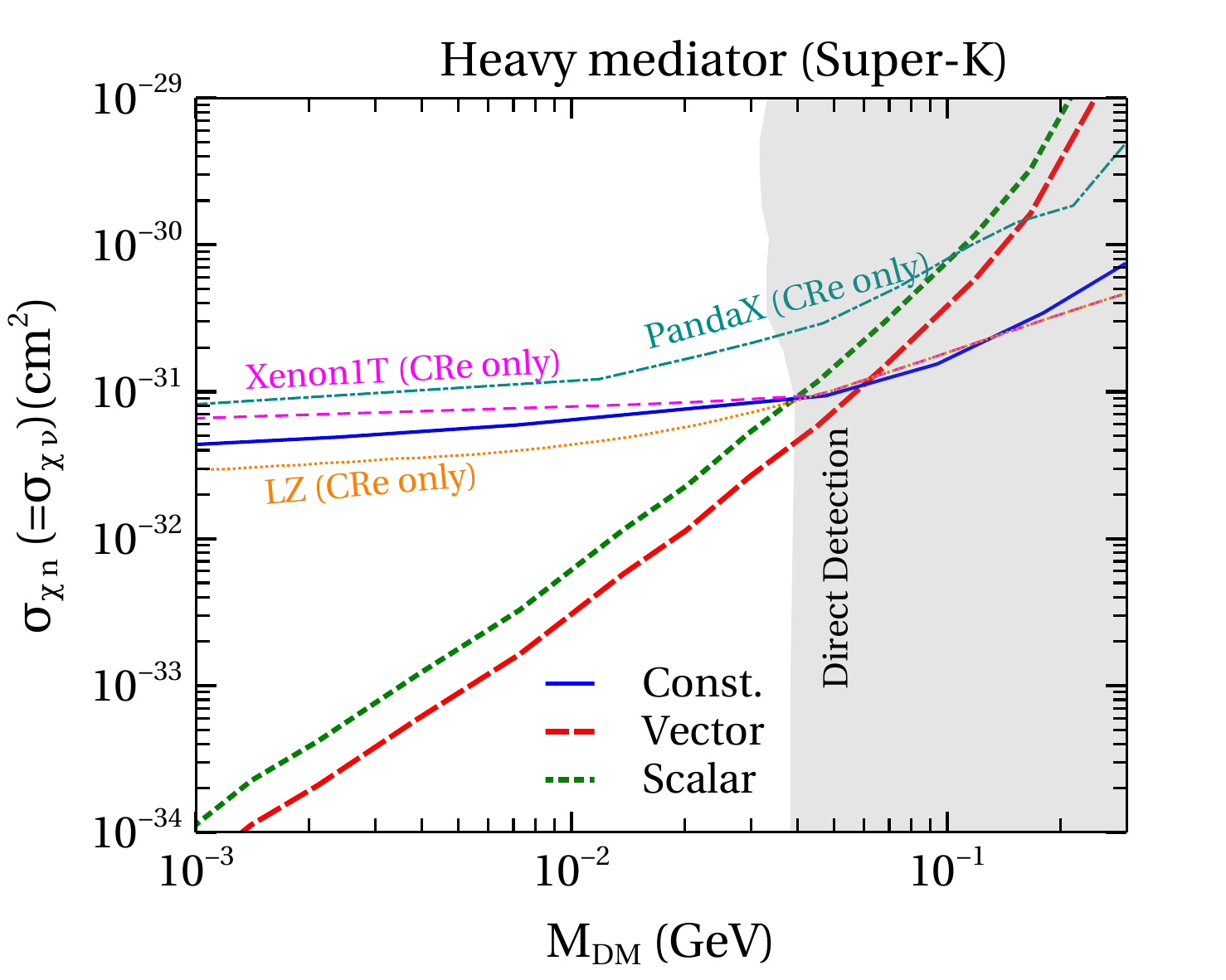}}
\caption{ Constraints on $M_{DM}$ vs. $\sigma_{\chi n}(=\sigma_{\nu \chi})$ plane with $90\%$ C.L from (a) XENONnT  (b) Super-Kamiokande assuming heavy mediator assuming only constant cross-sections.}
\end{figure}

Now, we turn our attention to the nucleophilic case and show the constraints in the $M_{DM}$ vs. $\sigma_{\chi n}$ plane.
Once again, we consider both CRp and DSNB effects considering $\sigma_{\chi n}=\sigma_{\chi \nu}$.
The results for XENONnT and Super-K are shown in \cref{n1,n2}, respectively by the blue dashed dotted line.
We showcase the existing direct detection constraints on $\sigma_{\chi n}$ for galactic DM coming mainly from SuperCDMS \cite{SuperCDMS:2018mne}, DAMIC \cite{DAMIC:2016lrs}, NEWS-G \cite{NEWS-G:2017pxg}, CRESST \cite{CRESST:2019jnq}, CDEX \cite{CDEX:2019hzn}, XENON \cite{XENON:2019zpr}, EDELWEISS \cite{Armengaud_2022}, and DarkSide \cite{DarkSide:2022dhx}. The combined exclusion region is shown in gray colour.
Constraint from space based experiment XQC \cite{Mahdawi:2018euy} is also displayed.
We also portray the constraints for DM boosted only by CR nucleons from XENON1T \cite{Bringmann:2018cvk}, LZ \cite{Maity:2022exk}, and PandaX-II \cite{PandaX-II:2021kai}.
After including the DSNB effects, our obtained constraints are stronger than them.
After including the DSNB effects the constraints do not modify significantly for Super-K with low mass DM for nucleon scattering compared to the electron scattering.
The reason is simply the higher target mass for nucleon recoil.
Since Super-K has a threshold at $E_R\sim\mathcal{O}(1)$ MeV, the minimum kinetic energy required to generate recoil signal for $\mathcal{O}$(MeV) DM is around $T_\chi^{\rm min}\sim\mathcal{O}(100)$ MeV.
At this energy, the boosted DM flux is mostly dominated by CRp up-scattered components.
However, assuming DM boosted after scattering with CR particles and DSNB can lead to new constraints on the previously allowed parameter space for low threshold experiments like XENONnT.

\begin{figure}[!tbh]
\centering
\subfigure[\label{cn1}]{
\includegraphics[scale=0.32]{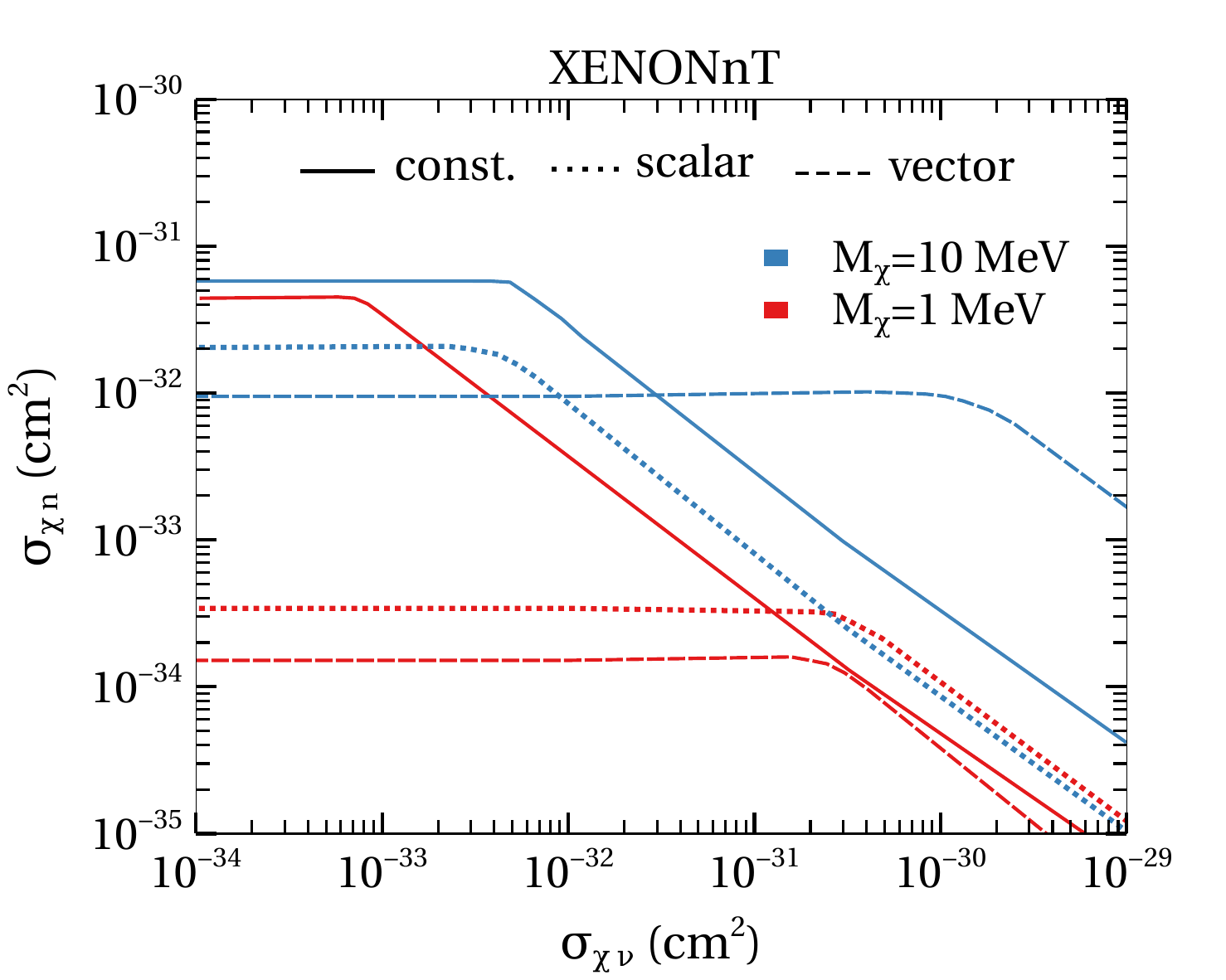}}
\subfigure[\label{cn2}]{
\includegraphics[scale=0.32]{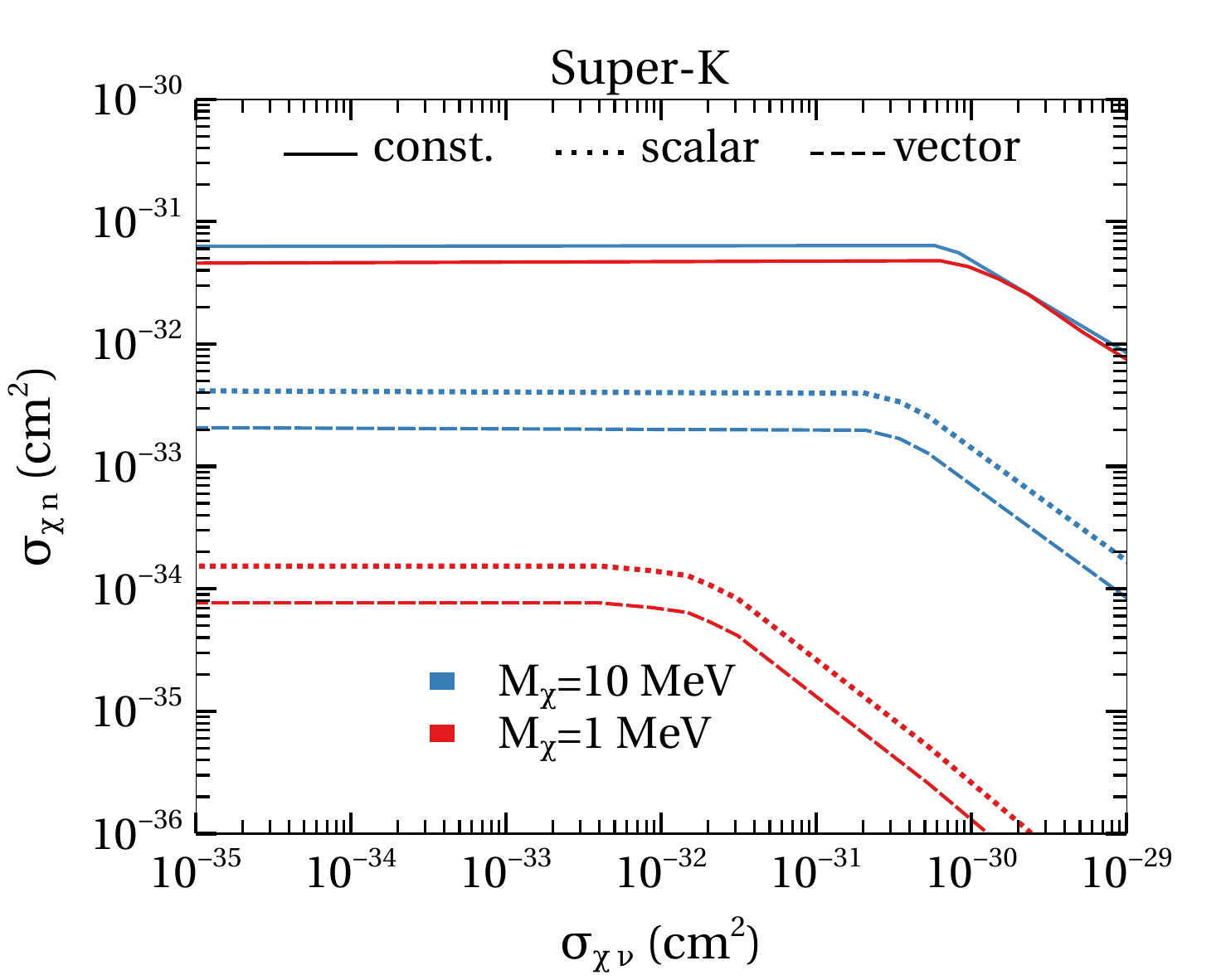}}
\caption{ Contour lines of $90\%$ C.L. constraints on the $\sigma_{\chi \nu}$ vs. $\sigma_{\chi n}$ plane for different values of $M_\chi$ from (a) XENONnT and (b) Super Kamiokande. The solid, dashed, and dotted lines represent constant, vector, and scalar cross-sections. The constraints corresponding to different $M_\chi\in \{1, 10\}$ MeV are shown by red and light blue colour, respectively. The regions above the contour lines are excluded.}
\label{fig:cont_n}
\end{figure}
In \cref{fig:cont_n}, we portray the $90\%$ C.L. constraints on the $\sigma_{\chi \nu}$ vs. $\sigma_{\chi n}$ plane for individual $M_\chi$ from XENONnT (\cref{c1}) and Super Kamiokande (\cref{c2}).
The constraints corresponding to different interactions: constant, vector-, and scalar- mediated are depicted by solid, dashed, and dotted lines.
The different values of $M_\chi \in \{1,10\}$ MeV are shown in red and light blue colour, respectively.
The behaviour of the contours can be easily understood as analogous to the electrophilic scenario from the discussion in the context of \cref{fig:cont_n}.

Throughout this work, we do not portray the possible upper bound in the exclusion regions which may come from the attenuation of DM particles in the earth's core.
Because the detectors are placed deep underground, a significant fraction of DM particles may scatter off nucleons (electrons) in the earth's crust and deplete their kinetic energy in this process.
The exact treatment of attenuation requires an exhaustive study and understanding of the composition of the earth's crust \cite{DeRomeri:2023ytt}, which is beyond the scope of this work.
However, a simple qualitative argument (see ref.\cite{Dent:2019krz, Bardhan:2022bdg}) reveals that the upper-bound from attenuation is quite high ($\sigma_{\chi n,e}\sim 10^{-28}~{\rm cm}^2$) and is already within the exclusion region of other existing constraints. 
Thus, despite the simplified approach, this work brings a new understanding of DM-neutrino interactions and indeed explores new parameter space.

\section{Conclusion}
\label{sec:conc}
DM particles with low masses have very limited kinetic energy, leading to extremely small energy transfers (in the sub-keV range) when they interact with SM particles. It is 
extremely difficult to detect such small energy signals because this requires detectors with exceptionally low-energy thresholds and high sensitivity. To achieve this, especially without background noise interfering, is a humongous task.
Despite extensive efforts, the nuclear interactions of sub-GeV DM remain largely 
unexplored in current direct detection experiments. While DM masses up to $\sim \mathcal {O}(1)$ MeV can be probed in existing direct detection experiments via DM-electron interactions, sub-MeV DM-electron interactions remain inaccessible. In recent years, boosted low-mass DM, specifically DM up-scattering by ultra high-energy CR flux, has attracted 
considerable attention and has been studied to 
impose limits on $\sigma_{\chi-e}$ or $\sigma_{\chi-n}$. 
This paper explores a minimal scenario with DM-$\nu$ interaction alongside either $\chi -e$ or $\chi-p$ interaction.
Apart from CR particles, theories also predict abundant DSNB flux in the universe and a non-zero 
$\sigma_{\chi \nu}$ allows DM to be boosted by DSNB as well.
This additional contribution helps to achieve stronger constraints on the cross-sections from experiments like XENONnT and Super-K.
We also consider the energy-dependent cross-sections like scalar and vector mediators which have not been considered to date, taking contributions from both CR and DSNB.
We show the constraints differ significantly for electrophilic DM assuming the energy-dependent cross-sections.
We perform a model-independent analysis of electrophilic and nucleophilic DM with a non-zero neutrino scattering cross-section ($\sigma _{\chi \nu}$). Super-K currently has the 
most stringent constraints in this scenario due to its larger target size. The combined effects of the DSNB and CR particles could potentially open up new parameter space in the low-mass region.
Apart from the DM-electron (nucleon) interactions this work also explores the possibility of probing the DM-neutrino interactions.
For a fixed DM mass, we vary $\sigma_{\chi \nu}$ and $\sigma_{\chi e}$ ($\sigma_{\chi n}$) to obtain the constraints in the cross-section plane.
With the combined effect of CR and DSNB, the proposed multiton experiments like DUNE, Hyper-Kamiokande, JUNO, etc. should probe even smaller cross-sections of DM in sub-MeV mass region and may provide further insights into low-mass DM-SM interactions.

\section*{Acknowledgement}
TG acknowledges Magnus Ehrnrooth Foundation for funding his doctoral studies. MH acknowledges support from the Research Council of Finland (grant \#342777). SJ is indebted to Anirban Majumdar for invaluable discussions and Vikramaditya Mondal for providing computational support. SJ also acknowledges CSIR, Government of India for financial support under the NET JRF fellowship scheme with
Award file No. 09/080(1172)/2020-EMR-I.

\bibliographystyle{utphys}
\bibliography{ref}

\end{document}